\documentclass{sig-alternate-05-2015}

\usepackage{graphicx}
\usepackage{balance}  
\usepackage[font=small,bf]{caption}
\usepackage[font=scriptsize]{subcaption}
\usepackage{cite}
\usepackage{enumitem}
\usepackage{cases}
\usepackage[all]{nowidow}

\usepackage{amsmath}
\usepackage{amssymb}

\usepackage{amsthm}
\usepackage[ruled,linesnumbered]{algorithm2e}
\usepackage{bbm}
\usepackage{bm}
\usepackage{makecell}
\usepackage{multirow}
\usepackage{pbox}
\usepackage{url}
\usepackage{xspace}
\usepackage[htt]{hyphenat}
\usepackage{alltt}
\usepackage{listings}
\usepackage{pifont}
\usepackage{newtxtext,newtxmath}

\usepackage{color}
\usepackage[normalem]{ulem}

\usepackage{hyperref}
\usepackage{cleveref}

\crefname{section}{Section}{Sections}
\Crefname{section}{Section}{Sections}
\crefname{appendix}{Appendix}{Appendices}
\Crefname{appendix}{Appendix}{Appendices}

\crefname{figure}{Figure}{Figures}
\Crefname{figure}{Figure}{Figures}
\crefname{equation}{Equation}{Equations}
\Crefname{equation}{Equation}{Equations}
\crefname{table}{Table}{Tables}
\Crefname{table}{Table}{Tables}
\crefname{lemma}{Lemma}{Lemmas}
\Crefname{lemma}{Lemma}{Lemmas}
\crefname{theorem}{Theorem}{Theorems}
\Crefname{theorem}{Theorem}{Theorems}

\crefname{algorithm}{Algorithm}{Algorithms}
\Crefname{algorithm}{Algorithm}{Algorithms}

\usepackage{tikz}
\usetikzlibrary{positioning}
\usetikzlibrary{calc}
\usetikzlibrary{decorations.markings}
\usepackage{pgfplots}
\usepgflibrary{shapes.geometric}
\usetikzlibrary{patterns,arrows}
\usetikzlibrary{arrows.meta}
\usepgfplotslibrary{fillbetween}

\usepackage{hyphenat}

\usepackage{xcolor}

\theoremstyle{definition}
\newtheorem{mydef}{Definition}

\newtheorem{theorem}{Theorem}
\newtheorem{lemma}{Lemma}

\DeclareMathOperator{\argmax}{Arg\,Max~~}
\DeclareMathOperator*{\erf}{erf}
\DeclareMathOperator*{\cov}{cov}

\newcommand{\etal}{{et al.\xspace}}
\newcommand{\eg}{{e.g.\xspace}}
\newcommand{\ie}{{i.e.\xspace}}

\makeatletter
\newcommand*{\etc}{%
    \@ifnextchar{.}%
        {etc}%
        {etc.\@\xspace}%
}
\makeatother

\newcommand{\ignore}[1]{}
\newcommand{\ph}[1]{\vspace{1mm} \noindent \textbf{#1}---}
\newcommand{\learn}{\textsc{Learner}\xspace}

\newcommand{\tans}{\bar{\theta}\xspace}
\newcommand{\rtans}{\bm{\bar{\theta}}\xspace}
\newcommand{\estans}{\theta\xspace}
\newcommand{\rawans}{\theta\xspace}

\newcommand{\estansvec}{\vec{\theta}\xspace}
\newcommand{\rans}{\bm{\theta}\xspace}

\newcommand{\mtuple}{\bm{t}\xspace}
\newcommand{\tcol}{a\xspace}

\newcommand{\filter}{F\xspace}

\newcommand{\tqcov}{\bar{\kappa}\xspace}

\newcommand{\qmean}{\mu\xspace}
\newcommand{\qmeanvec}{\vec{\mu}\xspace}

\newcommand{\impmean}{\widehat{\theta}_{n+1}\xspace}
\newcommand{\impstd}{\widehat{\beta}_{n+1}\xspace}
\newcommand{\mans}{\ddot{\theta}_{n+1}\xspace}
\newcommand{\mstd}{\ddot{\beta}_{n+1}\xspace}

\newcommand{\pr}{\mathrm{Pr}}

\newcommand{\dist}{\Delta}

\newcommand{\kvec}{\vec{k}\xspace}
\newcommand{\shift}{\bm{s}\xspace}
\newcommand{\shiftstd}{\eta\xspace}

\newcommand{\dbl}{Verdict\xspace}
\newcommand{\secdbl}{\protect{\fontsize{13}{12}\bfseries Verdict}\xspace}
\newcommand{\basel}{\textsc{Baseline2}\xspace}
\newcommand{\nol}{NoLearn\xspace}
\newcommand{\dblwo}{\textsc{{\dbl}NoAdjust}\xspace}
\newcommand{\dblw}{\textsc{{\dbl}Adjust}\xspace}
\newcommand{\Learning}{Learning\xspace}
\newcommand{\Inference}{Inference\xspace}

\newcommand{\vertica}{\texttt{Customer1}\xspace}
\newcommand{\tpch}{\texttt{TPC-H}\xspace}
\newcommand{\synthetic}{\texttt{Synthetic}\xspace}

\definecolor{orange}{rgb}{0.9,0.7,0.3}
\newcommand{\tofix}[1]{{\color{blue} #1}}
\newcommand{\rev}[2]{#2}
\newcommand{\afterrev}[1]{#1}

\newcommand{\barzan}[1]{{\textcolor{green}{[Barzan: #1]}}}
\newcommand{\young}[1]{{\textcolor{red}{#1}}}
\newcommand{\mike}[1]{}

\newlength{\neggap}
\renewcommand{\footnotesize}{\scriptsize}

\definecolor{tomato}{HTML}{F46524}

\definecolor{seagreen}{HTML}{334960}
\definecolor{seablue}{HTML}{B3B3FF}

\definecolor{RYB1}{RGB}{141, 211, 199}
\definecolor{RYB2}{RGB}{255, 255, 179}
\definecolor{RYB3}{RGB}{190, 186, 218}
\definecolor{RYB4}{RGB}{251, 128, 114}
\definecolor{RYB5}{RGB}{108, 157, 201}
\definecolor{RYB6}{RGB}{253, 180, 98}
\definecolor{RYB7}{RGB}{139, 0, 139}
\definecolor{RYB7}{RGB}{139, 0, 139}

\definecolor{darkteal}{HTML}{334960}
\definecolor{darkorange}{HTML}{F46524}
\definecolor{mygray}{HTML}{EBEDEF}
\definecolor{mPurple}{HTML}{9370DB}
\definecolor{mYellow}{HTML}{FFFF99}

\colorlet{blackdarkorange}{black!20!darkorange}
\colorlet{blackmPurple}{black!20!mPurple}
\colorlet{darkerteal}{black!90!darkteal}

\tikzset{bar7/.style={
    RYB7!50!black,fill=RYB7
}}

\tikzset{bar6/.style={
    RYB1!50!black,fill=RYB1
}}

\tikzset{bar5/.style={
    RYB2!50!black,fill=RYB2!50!black
}}

\tikzset{bar4/.style={
    RYB3!50!black,fill=RYB3
}}

\tikzset{bar3/.style={
    RYB6!50!black,fill=RYB6
}}

\tikzset{bar2/.style={
    RYB4!50!black,fill=RYB4
}}

\tikzset{bar1/.style={
    RYB5!50!black,fill=RYB5
}}

\renewenvironment{quote}
  {\list{}{\rightmargin=0.5cm \leftmargin=0.5cm}%
   \item\relax}
  {\endlist}

\begin{document}

\pgfplotsset {
    every axis/.append style={font=\scriptsize}
}

\pgfplotsset{
compat=newest,
every axis legend/.append style={font=\scriptsize, column sep=5pt},
/pgfplots/ybar legend/.style={
    /pgfplots/legend image code/.code={%
        \draw[##1,/tikz/.cd, bar width=6pt, yshift=-0.25em, bar shift=0pt, xshift=0.8em]
    plot coordinates {(0cm,0.8em)};}
}
}

%
\CopyrightYear{2017} 
\setcopyright{acmcopyright}
\conferenceinfo{SIGMOD'17,}{May 14-19, 2017, Chicago, IL, USA}
\isbn{978-1-4503-4197-4/17/05}\acmPrice{\$15.00}
\doi{http://dx.doi.org/10.1145/3035918.3064013}


\title{Database Learning:\\ Toward a Database that Becomes Smarter Every
  Time\titlenote{This manuscript is an extended report of the work published in
    ACM SIGMOD conference 2017.}}
%
%
%
%
%

\numberofauthors{1} 
%
\author{
%
%
\alignauthor
Yongjoo Park, Ahmad Shahab Tajik, Michael Cafarella, Barzan Mozafari\\
       \affaddr{University of Michigan, Ann Arbor}\\
       \email{\{pyongjoo, tajik, michjc, mozafari\}@umich.edu}
}

\maketitle
\begin{abstract}
  In today's databases, previous query answers \emph{rarely} benefit answering future queries.
  For the first time, to the best of our knowledge, we change this paradigm
   in an  approximate query processing (AQP) context.
  We make the  
  following observation: the answer to each query reveals some degree of \emph{knowledge} about the answer
  to another query because their answers stem from the same underlying distribution that has produced
  the entire dataset. Exploiting and refining this knowledge should allow us to answer
    queries more analytically, rather than by reading enormous amounts of raw
  data. Also, processing more queries should continuously enhance our knowledge of
  the underlying distribution, and hence lead to increasingly faster response times for future queries.

  We call this novel idea---learning from past query answers---\emph{Database
  Learning}.  We exploit \emph{the principle of maximum entropy} to produce
  answers, which are in expectation guaranteed to be more accurate than
  existing sample-based approximations. Empowered by this idea, we build a
  query engine on top of Spark SQL, called  \dbl.  We conduct extensive
  experiments on real-world query traces from a large customer of a major
  database vendor.  Our results demonstrate that \dbl supports
  73.7\% of these queries, speeding them up by up to 23.0$\times$ for the same
  accuracy level compared to existing AQP systems.

\end{abstract}





\section{Introduction}
\label{sec:intro}

In today's databases,
the answer to a previous query is rarely useful for speeding up new
queries.  Besides a few limited benefits (see \emph{Previous Approaches} below),
the
work (both I/O and computation) performed for answering past queries is often wasted
afterwards.
However, in
an approximate query processing context 
(e.g.,
\cite{mozafari_eurosys2013,surajit-optimized-stratified,aggr_oracle,easy_bound_bootstrap,online-agg,zeng2016iolap}),
one might be able to change this paradigm and reuse much of the previous work done by the database
system based on the following observation:

\begin{quote}
\emph{The  answer to each query reveals some \emph{fuzzy knowledge} about the answers to
	other queries, even if each query accesses a different subset of tuples and columns.}
\end{quote}

This is because the answers to different queries stem from the same (unknown) underlying
distribution that has generated the entire dataset.  In other words,
each answer reveals a piece of information about this underlying but
\textbf{unknown distribution}.  Note that having a concise statistical model of the
underlying data can have significant performance benefits.  In the ideal case,
if we had access to an incredibly precise model of the underlying data, we would
no longer have to access the data itself. In other words, we could answer
queries more efficiently by analytically evaluating them on our concise model,
which would mean reading and manipulating a few kilobytes of model parameters
rather than terabytes of raw data.  While we may never have a perfect model in
practice, even an imperfect model can be quite useful.  Instead of using the
entire data (or even a large sample of it), one can use a small sample of it to quickly produce a rough
approximate answer, which can then be calibrated and combined with the model to
obtain a more accurate approximate answer to the query.  \textbf{The more
precise our model, the less need for actual data, the smaller our sample, and
consequently, the faster our response time.} In particular, if we could somehow
continuously improve our model---say, by \emph{learning} a bit of information
from every query and its answer---we should be able to  \textbf{answer new queries
using increasingly smaller  portions of data, i.e., become smarter and faster as
we process more queries.}

We call the above goal \emph{Database Learning} (DBL), as it is reminiscent of the inferential goal of Machine Leaning (ML)
whereby  past
observations are used to improve future
predictions~\cite{ml-coined,ml-textbook-old,carlson2010toward}.
 Likewise, our goal in DBL is to enable a similar principle by \textbf{learning from past observations, but in a query
processing setting}. 
Specifically, in DBL, we plan to treat
approximate answers to past queries as observations, and  use them  to 
refine our
posterior knowledge of  the
underlying data, which in turn can be used to speed up future queries.


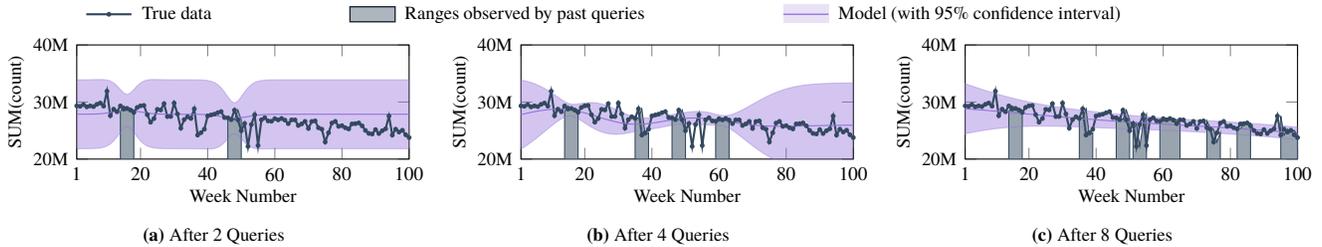
\begin{figure*}[t]

  \pgfplotsset{axisIntroFigure/.style={
          width=60mm,
          height=31mm,
          xmin=1,
          xmax=100,
          ymin=24000000,
          ymax=36000000,
          legend style={
            at={(0,1.10)},anchor=south west,column sep=1.5pt,
            draw=none,font=\scriptsize,fill=none,line width=.5pt,
            /tikz/every even column/.append style={column sep=50pt}
          },
          legend cell align=left,
          legend columns=3,
          clip=true,
          ytick={24000000, 30000000, 36000000},
          yticklabels={20M, 30M, 40M},
          xtick={1,20,40,60,80,100},
          ylabel=SUM(count),
          xlabel=Week Number,
          ylabel shift=-1mm,
          xlabel near ticks,
          xlabel shift=-1mm,
          scaled ticks=false,
  }}

  \pgfplotsset{curve/.style={%
          mark=none,
          opacity=1.0
  }}

  \tikzset{barstyle/.style={%
          opacity=0.5,
          darkteal,
          draw=black,
  }}

  \tikzset{legendnode/.style={%
          opacity=0.5,
          bar3,
          minimum width=6mm,
          minimum height=2.5mm,
          anchor=south west,
  }}

  \tikzset{queryarea/.style={%
      fill=darkteal!50!white,
      draw=darkteal,
  }}

  \tikzset{querynode/.style={%
      fill=darkteal!50!white,
      draw=darkteal,
      minimum width=5mm,
      minimum height=2.5mm,
      anchor=south west,
  }}

  \begin{subfigure}[b]{0.32\textwidth}
    \begin{tikzpicture}
      \begin{axis}[axisIntroFigure]

        \addplot[restrict x to domain=14:18,queryarea,forget plot]
        table  [x=x, y=g, col sep=tab] {data/intro_figure_model2_data.txt} \closedcycle;
        \addplot[restrict x to domain=46:50,queryarea,forget plot]
        table  [x=x, y=g, col sep=tab] {data/intro_figure_model2_data.txt} \closedcycle;

        \addplot[restrict x to domain=1:100,curve,darkteal,thick,
        mark=*,mark size=0.5pt]
        table [x=x, y=g, col sep=tab] {data/intro_figure_model3_data.txt};
        \addlegendentry{True data}

        \addlegendimage{area legend,fill=darkteal,draw=black!50!darkteal,fill opacity=0.4}
        \addlegendentry{Ranges observed by past queries}

        \addplot[restrict x to domain=1:100,curve,mPurple]
        table [x=x, y=x1, col sep=tab] {data/intro_figure_model3_data.txt};

        \addplot[restrict x to domain=1:100,name path=A,curve,mPurple,opacity=0.4,forget plot]
        table [x=x, y=x1t, col sep=tab] {data/intro_figure_model3_data.txt};
        \addplot[restrict x to domain=1:100,name path=B,curve,mPurple,opacity=0.4,forget plot]
        table [x=x, y=x1b, col sep=tab] {data/intro_figure_model3_data.txt};
        \addplot[mPurple,opacity=0.4,forget plot] fill between[of=A and B];


        \addlegendentry{Model (with 95\% confidence interval)}



      \end{axis}


      \node [legendnode,mPurple,opacity=0.2] at (9.39, 1.80) {};

    \end{tikzpicture}

    \caption{After 2 Queries}
  \end{subfigure}
  ~
  \begin{subfigure}[b]{0.32\textwidth}
    \begin{tikzpicture}
      \begin{axis}[axisIntroFigure]
        \addplot[restrict x to domain=14:18,queryarea,forget plot]
        table  [x=x, y=g, col sep=tab] {data/intro_figure_model2_data.txt} \closedcycle;
        \addplot[restrict x to domain=46:50,queryarea,forget plot]
        table  [x=x, y=g, col sep=tab] {data/intro_figure_model2_data.txt} \closedcycle;
        \addplot[restrict x to domain=35:39,queryarea,forget plot]
        table  [x=x, y=g, col sep=tab] {data/intro_figure_model2_data.txt} \closedcycle;
        \addplot[restrict x to domain=59:63,queryarea,forget plot]
        table  [x=x, y=g, col sep=tab] {data/intro_figure_model2_data.txt} \closedcycle;

        \addplot[restrict x to domain=0:100,curve,darkteal,thick,
        mark=*,mark size=0.5pt]
        table [x=x, y=g, col sep=tab] {data/intro_figure_model3_data.txt};
        \addplot[restrict x to domain=0:100,curve,mPurple]
        table [x=x, y=x2, col sep=tab] {data/intro_figure_model3_data.txt};

        \addplot[restrict x to domain=0:100,name path=A,curve,mPurple,opacity=0.4]
        table [x=x, y=x2t, col sep=tab] {data/intro_figure_model3_data.txt};
        \addplot[restrict x to domain=0:100,name path=B,curve,mPurple,opacity=0.4]
        table [x=x, y=x2b, col sep=tab] {data/intro_figure_model3_data.txt};
        \addplot[mPurple,opacity=0.4] fill between[of=A and B];

      \end{axis}
    \end{tikzpicture}

    \caption{After 4 Queries}
  \end{subfigure}
  ~
  \begin{subfigure}[b]{0.32\textwidth}
    \begin{tikzpicture}
      \begin{axis}[axisIntroFigure]
        \addplot[restrict x to domain=14:18,queryarea,forget plot]
        table  [x=x, y=g, col sep=tab] {data/intro_figure_model2_data.txt} \closedcycle;
        \addplot[restrict x to domain=46:50,queryarea,forget plot]
        table  [x=x, y=g, col sep=tab] {data/intro_figure_model2_data.txt} \closedcycle;
        \addplot[restrict x to domain=35:39,queryarea,forget plot]
        table  [x=x, y=g, col sep=tab] {data/intro_figure_model2_data.txt} \closedcycle;
        \addplot[restrict x to domain=59:65,queryarea,forget plot]
        table  [x=x, y=g, col sep=tab] {data/intro_figure_model2_data.txt} \closedcycle;

        \addplot[restrict x to domain=82:86,queryarea,forget plot]
        table  [x=x, y=g, col sep=tab] {data/intro_figure_model2_data.txt} \closedcycle;
        \addplot[restrict x to domain=51:55,queryarea,forget plot]
        table  [x=x, y=g, col sep=tab] {data/intro_figure_model2_data.txt} \closedcycle;
        \addplot[restrict x to domain=95:100,queryarea,forget plot]
        table  [x=x, y=g, col sep=tab] {data/intro_figure_model2_data.txt} \closedcycle;
        \addplot[restrict x to domain=73:77,queryarea,forget plot]
        table  [x=x, y=g, col sep=tab] {data/intro_figure_model2_data.txt} \closedcycle;

        \addplot[restrict x to domain=0:100,curve,darkteal,thick,
        mark=*,mark size=0.5pt]
        table [x=x, y=g, col sep=tab] {data/intro_figure_model3_data.txt};
        \addplot[restrict x to domain=0:100,curve,mPurple]
        table [x=x, y=x3, col sep=tab] {data/intro_figure_model3_data.txt};

        \addplot[restrict x to domain=0:100,name path=A,curve,mPurple,opacity=0.4]
        table [x=x, y=x3t, col sep=tab] {data/intro_figure_model3_data.txt};
        \addplot[restrict x to domain=0:100,name path=B,curve,mPurple,opacity=0.4]
        table [x=x, y=x3b, col sep=tab] {data/intro_figure_model3_data.txt};
        \addplot[mPurple,opacity=0.4] fill between[of=A and B];


      \end{axis}
    \end{tikzpicture}

    \caption{After 8 Queries}
  \end{subfigure}

  \caption{
    An example of how database learning might continuously refine its model as more queries are processed:
     after processing (a) 2 queries, (b) 4 queries, and (c) 8 queries.
    We could deliver more accurate answers if we
     combined this model with the approximate answers produced by traditional sampling techniques.
  }
  \label{fig:intro:model}
  \vspace{\neggap}
\end{figure*}

In \cref{fig:intro:model}, we visualize this idea using a 
real-world Twitter dataset~\cite{anderson2013brainwash,antenucci2016declarative}.
Here, DBL learns a model for the 
number of occurrences of certain word patterns (known as \emph{n-grams}, \eg, ``bought a
car'') in tweets.
\Cref{fig:intro:model}(a) shows this model (in purple) based on the answers to the
first two queries asking about the number of occurrences of these
  patterns, each over a different time range. Since the model is probabilistic, its 95\% confidence interval is
also shown (the shaded area around the best current estimate).
As shown in \cref{fig:intro:model}(b) and \cref{fig:intro:model}(c),
DBL further refines its model as more new queries are
answered.
This approach would allow a DBL-enabled query engine to provide increasingly more
accurate estimates, \emph{even for those ranges that have
never been accessed by previous queries}---this is possible because DBL
finds the most likely model  of the entire area that fits with the past query answers.
 The goal of this simplified example\footnote{In general, DBL does not make any \emph{a prior} assumptions regarding 
correlations (or smoothness) in the data; any correlations present in the data will be naturally
 revealed through analyzing the answers to past queries, in which case DBL
 automatically identifies and makes use of them.}
is to illustrate
 the possibility of (i) significantly faster response times by processing smaller samples of the data
for the same answer quality, or (ii) increasingly more accurate answers for the
same sample size and response time.



%

\ph{Challenges} 
To realize DBL's vision in practice, three key challenges  must be overcome.
First, there is a \emph{query generality} challenge.
DBL must be able to transform a wide class of SQL queries into
appropriate mathematical representations so that they can be
fed into statistical models and used for improving the accuracies of new queries.
Second, there is a \emph{data generality} challenge. 
To support arbitrary datasets, 
	DBL must not make any assumptions about the data distribution; 
the only valid
knowledge must come from past queries and their respective answers.
 Finally, there is an \emph{efficiency} challenge. DBL needs to strike a balance between 
	the computational complexity of its inference and its 
	ability to reduce the error of query answers. 
	In other words, DBL needs to be  both
effective and practical.


\ph{Previous Approaches}
In today's databases, the work performed for answering past queries is 
rarely  beneficial to new queries, except for the following cases:
\begin{enumerate}[leftmargin=5mm,noitemsep,nolistsep]
\item \textbf{View selection / Adaptive indexing}: In predictable workloads, columns and expressions commonly used  by past queries provide
hints on which indices~\cite{db-cracking,holistic-indexing,adaptive-indexing}
or materialized views~\cite{surajit-materialized} to build.
\item \textbf{Caching}: The recently accessed tuples might still be in memory 
when future queries access the same tuples.
\end{enumerate}

Both techniques, while beneficial, can only reuse previous work to a limited extent. 
Caching input tuples reduces I/O if the data size exceeds memory, but does not reuse query-specific computations.
Caching (intermediate) final results can reuse computation only if future (sub-)queries are \emph{identical} to those in the past.
While index selection techniques use the knowledge about which columns are commonly filtered on, an index per se does not allow for reusing computation
from one query to the next.  Adaptive indexing schemes (e.g., database cracking~\cite{db-cracking}) use each query to 
	incrementally refine an index to amortize the cost across queries.
	 However, there is
still an exponential number of possible column-sets that can be indexed.
Also, they do not reuse query-specific computations.
Finally,  materialized views
	are only beneficial when there is a strict structural compatibility---such as query containment or
equality---between past  and new queries~\cite{halevy2001answering}.


The fundamental difference between DBL and these traditional approaches lead to a few 
interesting
characteristics of DBL:
\begin{enumerate}[leftmargin=5mm,noitemsep,nolistsep]
  \item Since materialized views, indexing, and caching are for exact query processing, they are
    only effective when new queries touch previously accessed \rev{C10}{ranges}.
    On the
    contrary, DBL works in AQP settings; thus, DBL can benefit new queries even
    if they query ranges
    that were not touched by past queries.
 This is due to DBL's probabilistic model, which provides most likely extrapolation even for
    unobserved parts of data.
  \item Unlike indices and materialized views, DBL incurs \emph{little storage overhead} as it only
    retains the past $n$ aggregate queries and their answers.
    In
    contrast,  indices and materialized views grow in size as the data grows,
    while DBL's
    storage requirement remains \emph{oblivious to the data size} (see
    \cref{sec:related} for a detailed discussion).
\end{enumerate}

\ph{Our Approach}
Our vision of database learning (DBL)~\cite{mozafari_cidr2015} might be achieved in
different ways depending on the design decisions made in terms of
query generality, data generality, and efficiency.
In this paper, besides the introduction of the concept of DBL, 
we also provide a specific solution for achieving DBL, which we call \dbl to distinguish it  
from DBL as a general vision.

From a high-level, \dbl addresses the three challenges---query
generality, data generality, and efficiency---as follows.
First, \dbl supports SQL queries by decomposing them into
simpler atomic units,  called \emph{snippets}. 
The answer to a snippet is a single scalar value; thus, our belief on the answer to each
snippet can be expressed as a random variable, which can then be used in our mathematical model.
Second, to achieve data generality, \dbl employs a \emph{non-parametric} probabilistic model, which
is capable of representing arbitrary underlying distributions.
This model is based on a simple intuition: \emph{when two queries share
	some tuples in their aggregations, their answers must be correlated.} 
Our probabilistic model is a formal generalization of this idea using \emph{the principle of maximum
entropy}~\cite{skilling2006data}.
Third, to ensure  computational efficiency, we keep our probabilistic
model in an analytic form.  At query time, we only require  a matrix-vector
multiplication; thus, the overhead is negligible.

\ph{Contributions}
This paper makes the following contributions:
\begin{enumerate}[leftmargin=5mm,noitemsep,nolistsep]
  \item We introduce the novel  concept of \emph{database learning} (DBL). By
    learning from past query answers, DBL allows DBMS to continuously become smarter and faster
    at answering new  queries.
 \item We provide a concrete instantiation of DBL, called \dbl.
   \dbl's strategies  cover 63.6\% of
   TPC-H queries and 73.7\% of a real-world query trace from a 
   leading vendor of analytical DBMS.
   Formally, we also prove that \dbl's expected errors are never larger
   than those of existing AQP techniques.
 \item We integrate \dbl on top of Spark SQL, and conduct
   experiments using both benchmark and real-world query traces. 
    \dbl delivers up to 23$\times$ speedup and 90\% error reduction compared to AQP engines
    that do not use DBL. 
\end{enumerate}

The rest of this paper is organized as follows.  
\Cref{sec:overview} overviews \dbl's 
workflow, supported query types, and internal query representations.
 \Cref{sec:infer,sec:model} describe the internals of \dbl in
 detail, and \cref{sec:benefit} presents \dbl's formal guarantees.
\Cref{sec:summary} summarizes \dbl's online and offline processes, and
 \Cref{sec:overview:deployment} discusses \dbl's deployment scenarios.
 \cref{sec:exp} reports our empirical results.
 \Cref{sec:related} discusses related work, and \cref{sec:con}
 concludes the paper with future work.




\begin{figure}[t]
  \centering
  \begin{tikzpicture}[y=8.0mm]
    \tikzset{block/.style={
        draw=black,minimum height=8mm,minimum width=16mm,
        font=\scriptsize,
        align=center,
    }}

    \tikzset{figtext/.style={
        font=\scriptsize\tt,
        align=center,
    }}

    \tikzset{labeltext/.style={
        text=RYB6!50!black,
        fill=none,
        draw=none,
        align=left,
        font=\scriptsize\bf,
    }}

    \tikzset{onlineFlow/.style={
        RYB6!80!black,
        ultra thick,
    }}

    \tikzset{offlineFlow/.style={
        black!20!darkteal,
        ultra thick,
        dashed,
    }}

    \draw[bar3,opacity=0.2,rounded corners=2mm] (-1.9,-3.3) rectangle (2.0,1.7);  
    \draw[draw=black!50!darkteal,fill=darkteal,
    opacity=0.2,rounded corners=2mm] (4,-2.7) rectangle (6.5,0);  
    \node[labeltext] at (0,1.5) {Runtime query processing};
    \node[labeltext,text=black!50!darkteal] at (5.25,-0.4) {Post-query\\ processing};

    \node[figtext] (q) at (0,1) {SQL query};
    \node[block,font=\scriptsize]   (a) at (0,0) {AQP\\ Engine};
    \node[figtext] (b) at (0,-1) {(raw ans, raw err)};
    \node[block]   (i) at (0,-2) {\Inference};
    \node[figtext] (d) at (0,-3) {(improved ans, improved err)};
    \node[font=\bf\scriptsize,align=center] (sum) at (3,-1) {Query\\ Synopsis};
    \node[font=\bf\scriptsize,align=center] (m) at (3,-2) {Model};

    \draw[->,onlineFlow] (q) -- (a);
    \draw[->,onlineFlow] (a) -- ($(b.north)+(0,-0.1)$);
    \draw[->,onlineFlow] ($(b.south)+(0,0.1)$) -- (i);
    \draw[->,onlineFlow] (i) -- ($(d.north)+(0,-0.1)$);
    \draw[->,onlineFlow] ($(sum.south west)$) -- (i);
    \draw[->,onlineFlow] (m) -- (i);

    \coordinate (sumanchor) at (3,-1);
    \draw[->,offlineFlow] (b) -- (sum);

    \node[block] (learn) at (5.5,-2) {\Learning};
    \coordinate (sum_learn) at (5.5,-1);
    \draw[-,offlineFlow] (sum) -- (sum_learn);
    \draw[->,offlineFlow] (sum_learn) -- (learn);
    \draw[->,offlineFlow] (learn) -- (m);

    \coordinate (onlinelegend) at (2.5, 1.2);
    \coordinate (onlinelegend2) at ($ (onlinelegend) + (0.3,0) $);
    \draw[->,onlineFlow] (onlinelegend) -- (onlinelegend2);
    \node[font=\scriptsize,anchor=west] at ($ (onlinelegend2) + (0.1,0) $)
    {Runtime dataflow};

    \coordinate (offlinelegend) at ($ (onlinelegend) + (0, -0.5) $);
    \coordinate (offlinelegend2) at ($ (offlinelegend) + (0.3,0) $);
    \draw[->,offlineFlow] (offlinelegend) -- (offlinelegend2);
    \node[font=\scriptsize,anchor=west] at ($ (offlinelegend2) + (0.1,0) $)
    {Post-query dataflow};

  \end{tikzpicture}





  \caption{Workflow in \dbl. At query time, the \Inference module
	uses   the \emph{Query Synopsis} and the \emph{Model}
    to improve the query answer and error computed by the underlying AQP engine (\ie, \emph{raw
    answer/error}) before returning them to the user. 
    Each time a query is processed,
    	the raw answer and error are added to the \emph{Query Synopsis}.
    The \Learning module uses this updated \emph{Query Synopsis} to refine the current \emph{Model} accordingly.
  }
  \label{fig:overview:workflow}
  \vspace{\neggap}
\end{figure}
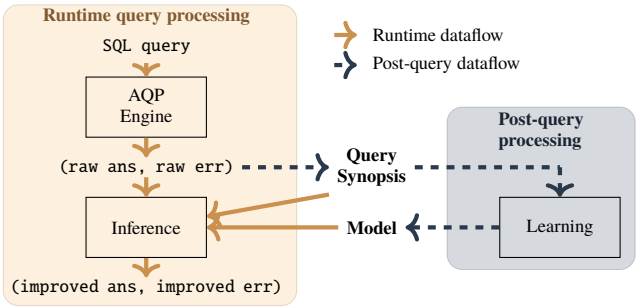

\section{Verdict Overview}
\label{sec:overview}

In this section, we overview the system we have built based on database learning
(DBL), called \dbl.
\Cref{sec:workflow} explains \dbl's architecture and overall workflow.
\Cref{sec:overview:query} presents the class of SQL queries currently supported by \dbl.
\Cref{sec:overview:internal} introduces
\dbl's query representation.
\Cref{sec:overview:infer} describes the intuition behind \dbl's inference.
Lastly, \cref{sec:overview:limit} discusses the limitations of \dbl's approach.

\subsection{Architecture and Workflow}
\label{sec:workflow}


\dbl consists of a \emph{query
synopsis}, a \emph{model},
and three processing modules: an
\emph{inference module}, a \emph{learning module},
and an off-the-shelf
\emph{approximate query processing (AQP) engine}.
Figure~\ref{fig:overview:workflow} depicts the connection between these components.


We begin by defining \emph{query snippets}, which serve as the basic units of
inference in \dbl.

 
\begin{mydef} \textbf{(Query Snippet)}
  A query snippet is a \emph{supported} SQL query whose answer is a
    \emph{single scalar value}, where supported queries are formally defined in
    \cref{sec:overview:query}.
\end{mydef}

\Cref{sec:overview:internal} describes how a supported query (whose answer may be a set)
is decomposed into possibly multiple query snippets.
For simplicity, and without loss of generality, 
here we assume that every incoming query is a query
snippet.

For the $i$-th query snippet $q_i$, the AQP engine's answer includes a pair of an
approximate answer
$\rawans_i$ and a corresponding expected error $\beta_i$. 
$\rawans_i$ and $\beta_i$ are formally defined in \cref{sec:infer:prob}, 
and are produced by most AQP
systems~\cite{online-agg,g-ola,online-agg-mr1,zeng2016iolap,mozafari_eurosys2013,mozafari_sigmod2014_abm}.
Now we can formally define the first key component of our system, the \emph{query synopsis}.

\begin{mydef} \textbf{(Query Synopsis)}
  Let $n$ be the number of query snippets processed thus far by the
  AQP engine. The query synopsis $Q_n$ is defined as the following set:
   $\{(q_i, \rawans_i, \beta_i) \mid i = 1, \ldots, n \}$.
\end{mydef}

We call the query snippets in the query synopsis \emph{past snippets}, and call the
$(n+1)$-th query snippet the \emph{new snippet}.

The second key component is the \emph{model},
 which represents \dbl's
statistical understanding of the underlying data. The model
is trained
on the query synopsis, and is updated every time a query is added to
the synopsis (\cref{sec:model}).

The query-time workflow of \dbl is as follows. Given an incoming query snippet
$q_{n+1}$, \dbl invokes the
AQP engine to compute a raw answer $\rawans_{n+1}$ and a raw error $\beta_{n+1}$.
Then, \dbl combines this raw answer/error and the previously
computed model to \emph{infer} an \emph{improved answer} $\impmean$ and an associated
expected error $\impstd$, called \emph{improved error}.
\Cref{thm:accuracy} shows that the improved error is never larger than
the raw error. After returning the improved answer and the improved error to the user, $(q_{n+1},
\rawans_{n+1}$, $\beta_{n+1})$ is added to the query synopsis.

\begin{table}[t]
  \centering
  \small

  \setlength\tabcolsep{4pt}
  \begin{tabular}{l p{5.8cm}}
    \hline
    \textbf{Term}       & \textbf{Definition} \\ \Xhline{2\arrayrulewidth}
    \textbf{raw answer}          & answer computed by the AQP engine \\[0.5mm]
    \textbf{raw error}           & expected error for raw answer  \\[0.5mm]
    \textbf{improved answer}     & answer updated by \dbl \\[0.5mm]
    \textbf{improved error}      & expected error for improved answer (by \dbl) \\[0.5mm]
    \textbf{past snippet}        & supported query snippet processed in the past \\[0.5mm]
    \textbf{new snippet}         & incoming query snippeet \\[0.5mm]
    \hline
  \end{tabular}

  \vspace{-2mm}
  \caption{Terminology.}
  \label{tab:terminology}
  \vspace{\neggap}
\end{table}


A key objective in \dbl's design is to treat the underlying AQP engine as a black box.
 This allows \dbl to be used 
 	with many of the existing engines without requiring any modifications.
From the  user's perspective, 
the benefit of using \dbl (compared to using the AQP engine alone)
	is the error reduction and speedup, or only the error reduction, depending on the type of AQP engine used
(\cref{sec:overview:deployment}).

Lastly, \dbl does not modify non-aggregate expressions or unsupported queries, \ie,
	it simply returns their raw answers/errors to the user.
\Cref{tab:terminology}
summarizes the terminology defined above. In \cref{sec:infer}, we will recap the mathematical
notations defined above.

\subsection{Supported Queries}
\label{sec:overview:query}

\lstset{aboveskip=0pt,belowskip=0pt}



\dbl supports aggregate queries that are flat (i.e., no derived tables or sub-queries)
with the following conditions:
\begin{enumerate}[leftmargin=5mm,noitemsep,nolistsep]
  \item \textbf{Aggregates}. Any number of \texttt{SUM}, \texttt{COUNT}, or
    \texttt{AVG} aggregates can appear in the \texttt{select} clause.
    The arguments to these aggregates can also be a \emph{derived
    attribute}.
    
  \item \textbf{Joins}.
    \rev{C14}{\dbl supports foreign-key joins between a fact
    table\footnote{Data warehouses typically record measurements (\eg, sales)
    into fact tables and  normalize commonly appearing attributes (\eg, seller
    information) into dimension tables~\cite{silberschatz1997database}.}
    and any number of dimension tables, exploiting the fact that this type of join does not
    introduce a sampling bias~\cite{join_synopses}.
  }
    For simplicity, our discussion in this paper is based on a denormalized table.

    \item \textbf{Selections}. \dbl currently supports equality and inequality
      comparisons for categorical and numeric
      attributes (including the \texttt{in} operator). 
      Currently, \dbl does not support disjunctions and
      textual filters (e.g.,  \texttt{like
      '\%Apple\%'}) in the \texttt{where} clause.
    
    \item \textbf{Grouping}. \texttt{groupby} clauses are supported for both stored and
      derived attributes. The query may also include a \texttt{having} clause.
        Note that the underlying AQP engine may affect the cardinality of the result set  depending on the \texttt{having} 
        clause (\eg, subset/superset error).
        \dbl simply operates on the result set returned by the AQP engine.
\end{enumerate}

\ph{Nested Query Support} Although \dbl does not directly support nested
queries, many queries can be flattened using
joins~\cite{query_flatten}
or by creating intermediate
views for sub-queries~\cite{halevy2001answering}.
In fact, this is the process used by Hive for supporting the
nested queries of the TPC-H benchmark~\cite{tpchonhive2009}.
We are currently working to automatically process nested queries and to expand
the class of supported queries (see \cref{sec:con}).


\ph{Unsupported Queries}
	Each query, upon its arrival,  is inspected by  \dbl's query type checker to determine whether it is supported, 
		and if not,  \dbl  bypasses the
    \Inference module and simply returns the raw answer to the user.
	The overhead of the query type
  checker is negligible (\cref{sec:exp:overhead}) compared to the runtime of the AQP engine;
thus, \dbl does not incur any noticeable runtime overhead, even when a query is not supported.

Only supported queries are stored in \dbl's
query synopsis and used to improve the accuracy of answers to future supported
queries.
That is, the class of queries that can be improved is equivalent to
the class of  queries that can be used to improve other queries.


\subsection{Internal Representation}
\label{sec:overview:internal}

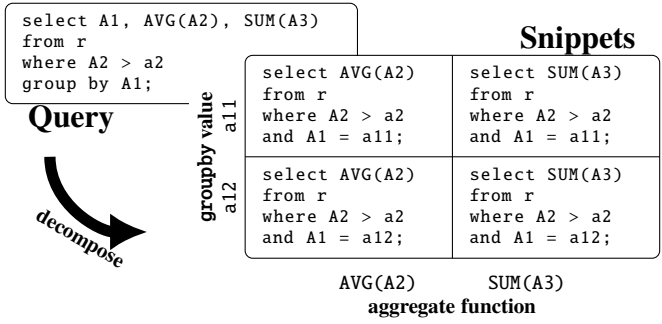
\begin{figure}[t]

  \tikzset{sqlbox/.style={
    fill=white,
  }}

  \centering
  \begin{tikzpicture}
    \node[sqlbox,draw=black,rounded corners=1mm] (q) at (0,0) {
      \begin{minipage}{0.53\linewidth}
        \begin{lstlisting}
select A1, AVG(A2), SUM(A3)
from r
where A2 > a2
group by A1;
        \end{lstlisting}
      \end{minipage}
    };

    \node[draw=none,fill=white,minimum width=5mm,minimum height=5mm] at ($(q.south)+(1,0)$) {};

    \node[sqlbox,anchor=west] (a) at ($(q.east)+(-1.5,-0.7)$) {
      \begin{minipage}{0.3\linewidth}
        \begin{lstlisting}
select AVG(A2)
from r
where A2 > a2
and A1 = a11;
        \end{lstlisting}
      \end{minipage}
    };

    \node[sqlbox,anchor=west] (b) at (a.east) {
      \begin{minipage}{0.3\linewidth}
        \begin{lstlisting}
select SUM(A3)
from r
where A2 > a2
and A1 = a11;
        \end{lstlisting}
      \end{minipage}
    };

    \node[sqlbox,anchor=north] (c) at (a.south) {
      \begin{minipage}{0.3\linewidth}
        \begin{lstlisting}
select AVG(A2)
from r
where A2 > a2
and A1 = a12;
        \end{lstlisting}
      \end{minipage}
    };

    \node[sqlbox,anchor=west] (d) at (c.east) {
      \begin{minipage}{0.3\linewidth}
        \begin{lstlisting}
select SUM(A3)
from r
where A2 > a2
and A1 = a12;
        \end{lstlisting}
      \end{minipage}
    };

    \draw[rounded corners=1mm] (a.north west) -- (b.north east) -- (d.south
    east) -- (c.south west) -- cycle;
    \draw (a.south west) -- (b.south east);
    \draw (a.north east) -- (c.south east);


    \node[rotate=90,align=center,fill=white,font=\small] at ($(a.south west)+(-0.38,0)$)
    {\bf \texttt{groupby} value \\ \texttt{a12} \hspace{5mm} \texttt{a11}};

    \node[align=center,fill=none,font=\small] at ($(c.south east)+(0,-0.5)$)
    {\texttt{AVG(A2)} \hspace{8mm} \texttt{SUM(A3)}\\
    \bf aggregate function};

    \node[anchor=west,font=\bf\large] at ($(q.south west)+(0.2,-0.2)$) {Query};
    \node[anchor=east,font=\bf\large] at ($(b.north east)+(-0.2,0.2)$) {Snippets};

    \coordinate (t) at ($(q.south west)+(0.6,-0.7)$);
    \draw[-{Latex[length=5mm,width=5mm]},line width=1.5mm]
    (t) to[out=270,in=180] ($(t)+(1.3,-1)$);

    \node[rotate=-30,font=\bf\small] at ($(q.south west)+(1.0,-1.8)$) {decompose};
  \end{tikzpicture}

  \caption{Example of a query's decomposition into multiple snippets.  }
  \label{fig:overview:snippet}
\end{figure}

\ph{Decomposing Queries into Snippets}
As mentioned in \cref{sec:workflow}, 
each supported query is broken into (possibly) multiple \emph{query snippets} before being added to the query synopsis.
Conceptually, each snippet corresponds to a supported SQL query with a single
aggregate function, with no other projected columns in its \texttt{select} clause, and with no \texttt{groupby} 
clause; thus, the answer 
to each snippet is a single scalar value.
A SQL query with
multiple aggregate functions (\eg, \texttt{AVG(A2)}, \texttt{SUM(A3)}) or a \texttt{groupby} clause is converted to
a set of multiple snippets 
for all combinations of each aggregate function and each \texttt{groupby}
column value. As shown in the example of \cref{fig:overview:snippet},  
each \texttt{groupby} column value is added as an
equality predicate in the \texttt{where} clause.
The number of generated snippets can be extremely large, \eg, if a
\texttt{groupby} clause includes a primary key. To ensure that the number of
snippets added per each query is bounded, \dbl only generates snippets
for $N^{max}$ (1,000 by default) groups in the answer set. \rev{C2.1b}{\dbl
computes improved answers only for those snippets
in order to bound the computational overhead.}\footnote{\rev{C16}{Dynamically adjusting
 the value of $N^{max}$ (e.g., based on available resources and workload characteristics)
makes an interesting direction for future work.}}
For each aggregate function $g$, the query synopsis retains a maximum of  $C_g$ query snippets by  
following a  least recently used snippet replacement policy (by default,
$C_g$$=$$2,000$).  
This improves the efficiency of the inference process, while maintaining an
accurate model based on the recently processed snippet answers.




\ph{Aggregate Computation}
\dbl uses two aggregate functions to perform its internal computations:
\texttt{AVG($A_k$)} and \texttt{FREQ(*)}.
 As stated earlier, the attribute $A_k$  can be either a stored attribute (\eg,
\texttt{revenue}) or a derived one (\eg, \texttt{revenue * discount}).
At runtime, \dbl combines these two types of aggregates to
compute
its supported aggregate functions as follows:
\begin{itemize}[itemsep=1pt,nolistsep]
  \item \texttt{AVG($A_k$)} = \texttt{AVG($A_k$)}
  \item \texttt{COUNT(*)} = round(\texttt{FREQ(*)} $\times$ (table cardinality))
  \item \texttt{SUM($A_k$)} = \texttt{AVG($A_k$)} $\times$ \texttt{COUNT(*)}
\end{itemize}

\subsection{Why and When Verdict Offers Benefit}
\label{sec:overview:infer}

In this section, we provide the high level intuition behind 
\dbl's approach to improving the quality of new snippet answers.
\dbl exploits potential correlations between snippet answers to infer the
answer of a new snippet. Let $S_i$ and $S_j$ be multisets of attribute
values such that, when aggregated, they output exact answers to queries $q_i$ and
$q_j$, respectively. 
Then, the answers to $q_i$ and $q_j$
are correlated, if:
\begin{enumerate}[nolistsep,itemsep=5pt,wide,labelwidth=!,labelindent=0pt]
  \item \textbf{$S_i$ and $S_j$ include common values.}
    $S_i \cap S_j \ne \phi$ implies the existence of correlation between the
    two snippet answers.
    For instance, computing the average revenue of the years 2014 and 2015 and
    the average revenue of the years 2015 and 2016 will be correlated since
    these averages include some common values (here, the 2015 revenue).
    In the TPC-H benchmark, 12 out of the 14 supported queries
 	 share common values in their
    aggregations.
  \item \textbf{$S_i$ and $S_j$ include correlated values.}
    For instance, the average prices of a stock over two consecutive days are
      likely to be similar even though they do not share common values.
      When the compared days are farther apart, the similarity in their average
      stock prices might be lower.
      \dbl captures the likelihood of such attribute value similarities
      using a statistical measure called
    \emph{inter-tuple covariance}, which will be formally defined in
    \cref{sec:data_stat}. In the presence of non-zero inter-tuple covariances,
    the answers to $q_i$ and $q_j$ could be correlated even when $S_i \cap S_j \ne \phi$.
 In practice, most real-life datasets
		tend to have non-zero  inter-tuple
    covariances, \ie, correlated attribute values (see \cref{sec:covariance:study} for an empirical study).
\end{enumerate}

\dbl formally captures the correlations between pairs of snippets 
using a probabilistic distribution function.
At query time, this probabilistic distribution function is used to infer
	the most likely answer to the new snippet given the answers to past snippets.

\subsection{Limitations}
\label{sec:overview:limit}


\dbl's model is the most likely explanation of the underlying distribution given the limited information
stored in the query synopsis. 
Consequently, when a new snippet involves tuples that have never been
accessed by past snippets, it is possible that \dbl's model might incorrectly represent the
underlying distribution, and return incorrect error bounds.
To guard against this limitation, 
 	\dbl always validates its model-based answer against the (model-free)
		answer of the AQP engine. 
We present this model validation step in \cref{sec:safeguard}.

%
%

Because \dbl relies on off-the-shelf  AQP engines for obtaining raw
answers and raw errors, it is naturally bound by the limitations of the underlying engine. For example,
it is known that sample-based engines are not apt at supporting arbitrary joins  or \texttt{MIN}/\texttt{MAX} aggregates.
Similarly, the validity of \dbl's error guarantees are contingent upon the
	validity of the AQP engine's raw errors. 
Fortunately, there are also off-the-shelf diagnostic
	techniques to verify the validity of such errors~\cite{mozafari_sigmod2014_diagnosis}.

\section{Inference}
\label{sec:infer}

In this section, we describe \dbl's inference process for computing an improved answer (and 
improved error) for the new snippet. 
\rev{C7}{\dbl's inference process follows the standard machine learning
arguments:
we can understand in part the true distribution by means of observations,
then we apply our understanding to predicting the unobserved.
To this end, \dbl applies well-established techniques, such as
the principle of maximum entropy and kernel-based estimations, to an AQP setting.}

To present our approach, we first
formally state our problem in \cref{sec:infer:prob}. A mathematical interpretation of the
problem and the overview on \dbl's approach is described in \cref{sec:infer:math}.
\Cref{sec:infer:prior,sec:infer:posterior} present the details of the \dbl's approach
to solving the problem.  \Cref{sec:infer:challenge} discusses
some challenges in applying \dbl's approach.


\subsection{Problem Statement}
\label{sec:infer:prob}

\begin{table}[t]
  \centering
  \small

  \begin{tabular}{l p{6.5cm}}
    \hline
    \textbf{Sym.}       & \textbf{Meaning} \\ \hline
    $q_{i}$             & $i$-th (supported) query snippet \\[1mm]
    $n+1$               & index number for a new snippet \\[1mm]
    $\rans_i$           & random variable representing our knowledge
                          of the raw answer to $q_i$ \\[1mm]
    $\rawans_i$         & (actual) raw answer computed by AQP engine for $q_i$  \\[1mm]
    $\beta_i$           & expected error associated with $\rawans_i$ \\[1mm]
    $\rtans_i$          & random variable representing our knowledge
                          of the \emph{exact} answer to $q_i$ \\[1mm]
    $\tans_i$           & exact answer to $q_i$ \\[1mm]
    $\impmean$          & improved answer to the new snippet \\[1mm]
    $\impstd$           & improved error to the new snippet \\[1mm]
    \hline
  \end{tabular}

  \vspace{-2mm}
\caption{Mathematical Notations.}

\label{tab:notations}
  \vspace{\neggap}
\end{table}

Let $r$ be a relation drawn from some unknown underlying distribution.
$r$ can be a join or Cartesian product of multiple tables.
Let $r$'s
attributes be $A_1, \ldots, A_m$, where $A_1, \ldots, A_l$ are the
\emph{dimension attributes} and $A_{l+1}, \ldots, A_m$ are the \emph{measure
attributes}.  Dimension attributes cannot appear
inside aggregate functions while measure attributes can.  Dimension attributes
can be numeric or categorical, but measure attributes are numeric.  Measure
attributes can also be \emph{derived attributes}.  \Cref{tab:notations} summarizes the notations we
defined earlier in \cref{sec:workflow}.

Given a query snippet $q_i$ on $r$, an AQP engine returns a raw answer
$\rawans_i$ along with an associated expected error $\beta_i$.
Formally,
$\beta_i^2$ is the expectation of the squared deviation of $\rawans_i$ from the (unknown) exact
answer $\tans_i$ to $q_i$.\footnote{Here, the expectation is made over $\rawans_i$ since the value
of $\rawans_i$ depends on samples.} $\beta_i$ and $\beta_j$ are independent if
$i \ne j$.

Suppose $n$ query snippets have been
processed, and therefore the query synopsis $Q_n$
contains the raw answers and raw errors for the past $n$ query snippets.
Without loss of generality, we assume all queries have the same aggregate
function $g$ on $A_k$ (\eg, \texttt{AVG($A_k$)}), where $A_k$ is  one of the measure
attributes. Our problem is then stated as follows: \emph{given $Q_n$ and ($\rawans_{n+1}$, $\beta_{n+1}$), compute the
most likely answer to $q_{n+1}$ with an associated expected error}.

In our discussion, for simplicity, we assume static data, \ie, the new
snippet is issued against the same data that has been used for answering past snippets in $Q_n$.
However, \dbl can also be extended to situations where the relations are
	subject to new data being added, \ie, each snippet is answered against a potentially different 
		version of the dataset.
    The generalization of \dbl under data updates is presented in \cref{sec:append}.

\subsection{Inference Overview}
\label{sec:infer:math}

In this section, we present our random variable interpretation of query answers
and a high-level overview of \dbl's inference process.

Our approach uses (probabilistic) random variables to represent
\emph{our knowledge of the query answers.}
The use of random variables here is a natural choice
as our knowledge itself of the query answers is
uncertain. Using random variables to represent degrees of belief
is a standard approach in Beyesian inference.
Specifically, we denote our knowledge of the raw answer
and the exact answer to the $i$-th query snippet
by random variables $\rans_i$ and $\rtans_i$, 
respectively.
At this step,
the only information available to us regarding $\rans_i$ and $\rtans_i$ is that $\estans_i$ is
an instance of $\rans_i$; no other assumptions are made.

Next, we represent   the relationship between the set of random variables
$\rans_1, \ldots, \rans_{n+1}, \rtans_{n+1}$
using a joint probability distribution function (pdf). 
 Note that the first $n+1$ random variables are for the raw answers to past $n$ snippets and the
 new snippet, and the last random variable is for the exact answer to the new snippet. 
We are interested in the relationship among those random
variables 
because our knowledge of the query answers is based on limited
information:
the raw answers computed by the AQP engine, whereas we aim to find the most likely value for the new snippet's exact answer. 
This joint pdf represents
\dbl's \emph{prior belief} over the query answers.
 We denote the joint
pdf by $f(\rans_1 =
\estans'_1, \ldots, \rans_{n+1} = \estans'_{n+1}, \rtans_{n+1} = \tans'_{n+1})$.
For brevity,
we also use $f(\estans'_1, \ldots, \estans'_{n+1}, \tans'_{n+1})$ when the meaning is clear from the context.
(Recall
 that $\estans_i$ refers to an actual raw answer from the AQP engine,  and
$\tans_{n+1}$ refers to the exact answer
to the new snippet.)
The joint pdf returns
the probability that the random variables $\rans_1, \ldots, \rans_{n+1}, \rtans_{n+1}$ takes a
particular combination of the values, \ie, $\estans'_1, \ldots, \estans'_{n+1}, \tans'_{n+1}$.
In \Cref{sec:infer:prior}, we discuss how to obtain this joint pdf from
some statistics available
on query answers.

Then, we compute the most likely value for the  new snippet's exact answer, namely the most likely value
for $\rtans_{n+1}$, by first conditionalizing the joint pdf on
the actual observations (\ie, raw answers) from the AQP engine, \ie,
$f(\rtans_{n+1} = \tans'_{n+1} \mid \rans_1 = \estans_1, \ldots, \rans_{n+1} = \estans_{n+1})$.
We then find the
value of $\tans'_{n+1}$ that maximizes the conditional pdf. 
We call this value the
\emph{model-based answer} and denote it by $\mans$.  
\Cref{sec:infer:posterior} provides more details of this process.
Finally, $\mans$ and its associated expected error $\mstd$ are returned as \dbl's improved answer and
improved error if they pass the model validation (described in \cref{sec:safeguard}). 
Otherwise, the (original) raw answer
	and error are taken as \dbl's improved answer and error,
  respectively. In other words, if the model validation fails, \dbl simply 
	returns the original raw results from the AQP engine without any improvements.

\subsection{Prior Belief}
\label{sec:infer:prior}

In this section, we describe how \dbl obtains a joint pdf $f(\estans'_1,
\ldots, \estans'_{n+1}, \tans'_{n+1})$ that represents its knowledge of the
underlying distribution.
The intuition behind \dbl's inference is to make use of possible correlations between pairs of query
answers. 
This section applies such statistical information of query answers (\ie, means,
covariances, and variances) for obtaining the most likely joint pdf.  Obtaining the query statistics
is described in \cref{sec:model}.



To obtain the joint pdf, \dbl relies on the principle of maximum
entropy (ME)~\cite{skilling2006data,berger1996maximum}, a simple but powerful statistical tool for
determining a pdf of random variables given some statistical
information available.
According to the ME principle, given some testable information on 
random variables associated with
a pdf in question, the pdf that best represents the current state of our knowledge is the
one that maximizes the following expression, called \emph{entropy}:
\begin{align}
h(f) = - \int
f(\estansvec) \cdot \log f(\estansvec) \; d\estansvec
\label{eq:entropy}
\end{align}
where $\estansvec = (\estans'_1, \ldots, \estans'_{n+1}, \tans'_{n+1})$.

Note that the joint pdf maximizing the above entropy differs depending on the
kinds of given testable information, \ie, query statistics in our context. For
instance, the maximum entropy pdf given means of random variables is different
from the maximum entropy pdf given means and (co)variances of random variables.
In fact, there are two conflicting considerations when applying this principle. 
On one hand, the resulting pdf can be computed more efficiently if the provided statistics are
simple or few,
	\ie, simple statistics reduce the computational complexity. 
On the other hand,  the resulting pdf can describe
the relationship among the random variables more accurately if richer statistics are
provided, \ie, the richer the statistics, the more accurate our improved answers.
  Therefore, we need to choose an appropriate degree of statistical information
  	to strike a balance between
	the computational efficiency of pdf evaluation and its accuracy in describing
		the relationship among query answers.

Among possible options, \dbl uses the first and the second order
statistics of the random variables, \ie, mean, variances, and covariances.  The use of second-order
statistics enables us to capture the relationship among the answers to
different query snippets, while the joint pdf that maximizes the entropy can be expressed in an
analytic form. The uses of analytic forms provides computational efficiency.
Specifically, the joint pdf that maximizes
the entropy while satisfying the given means, variances, and covariances is a multivariate normal
with the corresponding means, variances, and covariances~\cite{skilling2006data}.

\begingroup

\setlength{\abovedisplayskip}{2pt}
\setlength{\belowdisplayskip}{3pt}

\begin{lemma}
Let $\bm{\vec{\theta}}=(\rans_1, \ldots, \rans_{n+1}, \rtans_{n+1})^\intercal$ be a vector of $n$$+$$2$
	random variables  with
mean values $\qmeanvec = (\qmean_1, \ldots, \qmean_{n+1}, \bar{\mu}_{n+1})^\intercal$ and 
a $(n$$+$$2)$$\times$$(n$$+$$2)$  covariance matrix $\Sigma$ specifying their variances and pairwise covariances.
    The joint pdf $f$ over these random variables that maximizes $h(f)$ while satisfying the provided means, variances,
    and covariances  is the following function:
    \begin{align}
      f(\estansvec)
      = \frac{1}{\sqrt{(2 \pi)^{n+2} |\Sigma|}}
      \exp \left(
        -\frac{1}{2} (\estansvec - \qmeanvec)^\intercal \Sigma^{-1} (\estansvec - \qmeanvec)
      \right),
      \label{eq:prior_pdf}
    \end{align}
    \label{lemma:pdf}
    and this solution is unique.
\end{lemma}
\endgroup

In the following section, we describe how \dbl computes the most likely answer to the new snippet
using this joint pdf in \cref{eq:prior_pdf}. We call the most likely answer a \emph{model-based
answer}. In \cref{sec:safeguard}, this model-based answer is chosen as an improved answer if it
passes a model validation.
Finally, in \cref{sec:infer:challenge}, we discuss the challenges involved in obtaining $\qmeanvec$
and $\Sigma$, \ie, the query statistics required for deriving the joint pdf.


\subsection{Model-based Answer}
\label{sec:infer:posterior}



In the previous section, we formalized the relationship among query answers, namely
$(\rans_1, \ldots, \rans_{n+1}, \rtans_{n+1})$, using a joint pdf.
In this section, we exploit this joint pdf to infer the most likely answer to
the new snippet.
In other words, we aim to find the most likely value for $\rtans_{n+1}$ (the random variable representing
$q_{n+1}$'s exact answer),
given the observed values for $\rans_1, \ldots, \rans_{n+1}$, \ie, the raw answers from the AQP
engine. We call the most likely value a \emph{model-based answer} and its associated
expected error a \emph{model-based error}.
Mathematically, \dbl's model-based answer $\mans$ to $q_{n+1}$ can be expressed as:
\begin{align}
    \mans =  \underset{\tans'_{n+1}}{\argmax}
    f ( \tans'_{n+1} \mid \rans_1 = \rawans_1, \ldots, \rans_{n+1} =
    \rawans_{n+1}).
    \label{eq:impmean}
\end{align}
That is, $\mans$ is the value at which the conditional pdf has its maximum value. The conditional
pdf, $f(\tans'_{n+1} \mid \estans_1, \ldots, \estans_{n+1} )$, is obtained by conditioning the joint pdf obtained in \cref{sec:infer:prior} on the
observed values, \ie, raw answers to the past snippets and the new snippet.


Computing a conditional pdf may be a computationally expensive task. However, a conditional
pdf of a multivariate normal distribution is analytically computable; it is
another normal distribution. Specifically, the conditional
pdf in \cref{eq:impmean} is a normal distribution with the following mean
$\mu_c$ and variance $\sigma_c^2$~\cite{bishop2006pattern}:
\begin{align}
  \mu_c &= \bar{\mu}_{n+1} + \kvec_{n+1}^\intercal \Sigma_{n+1}^{-1}
  (\estansvec_{n+1} - \qmeanvec_{n+1})
  \label{eq:infer} \\
  \sigma_c^2 &= \tqcov^2 - \kvec_{n+1}^\intercal \Sigma_{n+1}^{-1} \kvec_{n+1}
    \label{eq:impstd}
\end{align}
where:
\begin{itemize}[itemsep=0pt]
\item $\kvec_{n+1}$ is a column vector of length $n+1$ whose $i$-th element is $(i,n+2)$-th entry of $\Sigma$;

\item $\Sigma_{n+1}$ is a $(n+1)\times(n+1)$ submatrix  of $\Sigma$ consisting
  of $\Sigma$'s first $n+1$ rows and columns;

\item $\estansvec_{n+1}$$=$$(\rawans_1, \ldots, \rawans_{n+1})^\intercal$;

\item $\qmeanvec_{n+1} = (\qmean_1, \ldots, \qmean_{n+1})^\intercal$; and

\item $\tqcov^2$ is the $(n+2,n+2)$-th entry of $\Sigma$
\end{itemize}
Since the mean of a normal distribution is the value at which the pdf
takes a maximum value, we take $\mu_c$   as our
model-based answer $\mans$. Likewise, 
the expectation of the squared deviation of the value
  $\tans'_{n+1}$, which  is distributed
according to the conditional pdf in \cref{eq:impmean}, from the model-based answer $\mans$
coincides with the variance $\sigma_c^2$ of the conditional pdf. Thus, we take $\sigma_c$ as 
our model-based error $\mstd$.




Computing each of \cref{eq:infer,eq:impstd} requires $O(n^3)$ time
complexity at query time. However, \dbl uses alternative
forms of these equations that require only $O(n^2)$ time complexity at query
time (\cref{sec:benefit}).
\rev{C13}{As a future work, we plan to employ 
inferential techniques with sub-linear time
complexity~\cite{lawrence2003fast,williams2000using} for a more sophisticated eviction policy for past queries.}

Note that, since the conditional pdf is a normal distribution, the error bound at confidence
$\delta$ is expressed as $\alpha_\delta \cdot \mstd$, where $\alpha_\delta$ is
a non-negative number such that a random number drawn from a standard normal distribution
	would fall within $(-\alpha_\delta, \alpha_\delta)$ with probability $\delta$. We call $\alpha_\delta$
the confidence interval multiplier for probability $\delta$. That is, 
the exact answer $\tans_{n+1}$ is within the range $(\mans - \alpha_\delta
\cdot \mstd, \; \mans + \alpha_\delta \cdot \mstd)$ with probability $\delta$,
according to \dbl's model.





\subsection{Key Challenges}
\label{sec:infer:challenge}

As mentioned in \cref{sec:infer:prior}, obtaining the joint pdf in Lemma~\ref{lemma:pdf}
(which
represents \dbl's prior belief on query answers) 
	requires the knowledge of means, variances, and covariances of the random
	variables $(\rans_1, \ldots, \rans_{n+1}, \rtans_{n+1})$.
However, acquiring these statistics is a non-trivial task for two
reasons. 
First, we have only observed one value for each of the random values
$\rans_1,\ldots,\rans_{n+1}$, namely $\rawans_1,$ $\ldots,$ $\rawans_{n+1}$.
Estimating variances and covariances of random variables from a single value is nearly impossible. 
Second, we do not have any observation for the last random variable
 $\rtans_{n+1}$ (recall that $\rtans_{n+1}$ represents our knowledge of
the exact answer to the new snippet, \ie, $\tans_{n+1}$).
In \cref{sec:model}, we present \dbl's approach to solving these
challenges.

	


\section{Estimating Query Statistics}
\label{sec:model}

As described in \cref{sec:infer}, \dbl expresses its prior belief
on the
relationship among query answers as a joint pdf over a set of random variables $(\rans_1, \ldots,$ $\rans_{n+1}, \rtans_{n+1})$. 
 In this process, we need to know the means, variances, and covariances of these random variables.



\dbl uses the arithmetic mean of the past query
answers for the mean of each random variable, 
$\rans_1, \ldots, \rans_{n+1}, \rtans_{n+1}$.
Note that
this only serves as a prior belief, and will be updated in the process of
conditioning the prior belief using the observed query answers.
In this section, without loss of generality, we
assume the mean of the past query answers is zero.

%

Thus, in the rest of this section, we focus on obtaining the variances and covariances of these random
variables, \rev{C2.1e}{which are the elements of the $(n+2) \times (n+2)$ covariance matrix $\Sigma$ in
  \cref{lemma:pdf} (thus, we can obtain the elements of the column vector
  $\kvec_{n+1}$ and the variance $\tqcov^2$ as well).
  Note that, due to the independence between expected errors, we have:
\begin{align}
  \begin{split}
    \cov(\rans_i, \rans_j) &= \cov(\rtans_i, \rtans_j) + \delta(i,j) \cdot \beta_i^2 \\
    \cov(\rans_i, \rtans_j) &= \cov(\rtans_i, \rtans_j)
  \end{split}
  \label{eq:query_cov2}
\end{align}
where $\delta(i,j)$ returns 1 if $i = j$ and 0 otherwise.
Thus, computing $\cov(\rtans_i, \rtans_j)$ is sufficient for obtaining $\Sigma$.
}

Computing $\cov(\rtans_i, \rtans_j)$ relies on a straightforward
observation: \emph{the covariance between two query snippet answers is computable using the covariances between the
attribute values involved in computing those answers.}
For instance, we can easily compute the covariance between (i) the average revenue of the years 2014 and 2015 and (ii) the
average revenue of the years 2015 and 2016, as long as we know the covariance between the
average revenues of every pair of days in 2014--2016.

\rev{C2.2}{In this work, we further extend the above observation. That is, if
we are able to compute the covariance between the average revenues at an infinitesimal
time $t$ and another infinitesimal time $t'$, we will be able to compute the
covariance between (i) the average revenue of 2014--2015 and (ii) the average
revenue of 2015--2016, by integrating the covariances between the revenues at infinitesimal
times over appropriate ranges. Here, the covariance between the average revenues
at two infinitesimal times $t$ and $t'$ is defined in terms of the underlying data
distribution that has generated the relation $r$, where the past query answers 
	help us discover the most-likely underlying distribution.}
The rest of this section formalizes this idea.

In \cref{sec:query_stat}, we present a decomposition of
the (co)variances between pairs of query snippet answers into \emph{inter-tuple covariance} terms.
Then, in \cref{sec:data_stat}, we describe how inter-tuple
covariances can be estimated analytically using parameterized functions.

\subsection{Covariance Decomposition}
\label{sec:query_stat}




To compute the variances and covariances between query snippet answers (\ie, $\rans_1,
\ldots, \rans_{n+1}, \rtans_{n+1}$), \dbl relies on our proposed \emph{inter-tuple covariances},
which express the statistical properties of the underlying distribution. Before
presenting the inter-tuple covariances, our discussion starts with the fact that
the answer to a supported snippet can be mathematically represented in terms of
the underlying distribution. This representation then naturally leads us to the
decomposition of the covariance
between query answers into smaller units, which we call inter-tuple
covariances.

Let $g$ be an aggregate function on attribute $A_k$, and $\mtuple = (a_1,
\ldots, a_l)$ be a vector of length $l$ comprised of the values for $r$'s
dimension attributes $A_1, \ldots, A_l$.
\rev{C1, C2.1a}{To help simplify the mathematical descriptions in this section,
we assume that all dimension attributes are numeric (not categorical),
and the selection predicates in queries may contain
range constraints on some of those dimension attributes.
Handling categorical attributes is a straightforward extension of this process
(see \cref{sec:discrete}).}

We define a continuous function
$\nu_g(\mtuple)$ for every aggregate function $g$ (\eg, \texttt{AVG($A_k$)},
\texttt{FREQ(*)}) such that, when integrated,
it produces answers to query snippets.
That is (omitting possible
normalization and weight terms for simplicity):
\begin{align}
  \rtans_i = \int_{\mtuple \in \filter_i} \, \nu_g(\mtuple) \; d\mtuple
  \label{eq:query_answer}
\end{align}
\rev{C2.1a}{Formally, $\filter_i$ is a
subset of the Cartesian product of the domains of the dimension
attributes, $A_1, \ldots, A_l$, such that $\mtuple \in \filter_i$ satisfies
the selection predicates of $q_i$. Let $(s_{i,k}, e_{i,k})$ be the range
constraint for $A_k$ specified in $q_i$. We set the range to 
$(\text{min}(A_k),~\text{max}(A_k))$ if no constraint is specified for $A_k$.
\dbl simply represents $\filter_i$ as the product of those $l$ per-attribute
ranges. Thus, the above \cref{eq:query_answer} can be expanded as:
\begin{align*}
  \rtans_i = \int_{s_{i,l}}^{e_{i,l}} \cdots \int_{s_{i,1}}^{e_{i,1}}
  \, \nu_g(\mtuple) \; d a_1 \cdots d a_l
\end{align*}
For brevity, we use the single integral representation using
$\filter_i$ unless the explicit expression is needed.}


Using \cref{eq:query_answer} and the linearity of covariance,
we can decompose $\cov(\rtans_i, \rtans_j)$ into:
\begin{align}
  \begin{split}
    \cov(\rtans_i, \rtans_j) &=
    \cov \left( \int_{\mtuple \in \filter_i} \nu_g(\mtuple) \; d\mtuple,
    \int_{\mtuple' \in \filter_j} \nu_g(\mtuple') \; d\mtuple' \right)  \\
    &= \int_{\mtuple \in \filter_i} \int_{\mtuple' \in \filter_j}
      \cov(\nu_g(\mtuple), \nu_g(\mtuple')) \; d\mtuple \; d\mtuple'
  \end{split}
  \label{eq:query_cov}
\end{align}

As a result, the covariance between query answers can be broken 
into an integration of the covariances between tuple-level function values, which we call
\emph{inter-tuple covariances}.


To use \cref{eq:query_cov}, we must be able to compute the inter-tuple
covariance terms.  However, computing these  inter-tuple covariances is
challenging, as we only have a single observation for each $\nu_g(\mtuple)$.
Moreover, even if we had a way to compute the inter-tuple covariance for
arbitrary $\mtuple$ and $\mtuple'$,
\rev{C2.1a}{the exact computation of \cref{eq:query_cov} would still require
an infinite number of
inter-tuple covariance computations, which would be infeasible.}
In the next section, we present an
efficient alternative for estimating these inter-tuple covariances.

\subsection{Analytic Inter-tuple Covariances}
\label{sec:data_stat}


To efficiently estimate the inter-tuple covariances, and thereby compute
\cref{eq:query_cov},
  we propose using \emph{analytical covariance functions}, a well-known technique in statistical literature for approximating 
  covariances~\cite{bishop2006pattern}. 
  In particular,
\dbl uses squared exponential covariance functions, 
which is capable of 
approximating any continuous target function  arbitrarily closely as
the number of observations (here, query answers) increases
~\cite{micchelli2006universal}.\footnote{
\rev{C1}{This property of the universal kernels is asymptotic (\ie, as the
number of observations goes to infinity).}}
\rev{C1}{Although the underlying distribution may not be a continuous
function, it is sufficient for us to obtain $\nu_g(\mtuple)$ such that, when
integrated (as in \cref{eq:query_answer}), produces the same values as
the integrations of the underlying distribution.}\footnote{\rev{C1}{The existence of such a continuous
function is implied by the kernel density estimation
technique~\cite{wasserman2006all}.}}
%
In our setting, 
the squared exponential covariance function $\rho_g(\mtuple, \mtuple')$
is defined as:
\begin{align}
  \cov(\nu_g(\mtuple), \nu_g(\mtuple'))
  \approx \rho_g(\mtuple, \mtuple')
  = \sigma_{g}^2 \cdot \prod_{k=1}^{l} \exp \left(
    -\frac{(\tcol_k - \tcol'_k)^2}{l_{g,k}^2} \right)
  \label{eq:tuple_cov}
\end{align}
Here, $l_{g,k}$ for $k$$=$$1\ldots l$ and $\sigma_g^2$ are tunable
\emph{correlation parameters} 
to be learned from past queries and their answers (\cref{sec:learn}).

Intuitively, when $\mtuple$ and $\mtuple'$ are similar, i.e.,
$(\tcol_k - \tcol'_k)^2$ is small for most $A_k$,
then  $\rho_g(\mtuple,
\mtuple')$ returns a larger value (closer to $\sigma_{g}^2$), indicating that the
expected values of $g$ for $\mtuple$ and $\mtuple'$ are highly correlated.  

\rev{C2.1c}{
  With the analytic covariance function above, the $\cov(\rtans_i, \rtans_j)$
  terms involving inter-tuple covariances can in turn be computed
  analytically. 
Note that \cref{eq:tuple_cov} involves the multiplication of $l$ terms, each
of which containing variables related to a single attribute. As a result, plugging
\cref{eq:tuple_cov} into \cref{eq:query_cov} yields:
\begin{align}
  \cov(\rtans_i, \rtans_j) &=
  \sigma_{g}^2
  \prod_{k=1}^{l}
    \int_{s_{i,k}}^{e_{i,k}}
    \int_{s_{j,k}}^{e_{j,k}} \,
    \exp \left( -\frac{(\tcol_k-\tcol'_k)^2}{l_{g,k}^2} \right) \,
    d a'_k a_k
  \label{eq:simple_cov}
\end{align}

The order of integrals are interchangeable, since the terms including no
integration variables can be regarded as constants (and thus can be factored out of the
integrals).
Note that the double-integral of an exponential function can also be computed
analytically (see \cref{sec:double_int}); thus,
\dbl can efficiently compute $\cov(\rtans_i, \rtans_j)$ in $O(l)$ times by
directly computing the integrals of inter-tuple covariances, without
explicitly computing individual inter-tuple covariances.}
\rev{C2.1e}{Finally, we can compose the $(n+2) \times (n+2)$ matrix
$\Sigma$ in \Cref{lemma:pdf} using \cref{eq:query_cov2}.}

\section{Formal Guarantees}
\label{sec:benefit}


%


Next, we formally show that the error bounds of \dbl's improved answers are never
larger than the error bounds of the AQP engine's raw answers.

\begingroup

\setlength{\abovedisplayskip}{2pt}
\setlength{\belowdisplayskip}{3pt}

\begin{theorem}
  Let \dbl's improved answer and improved error to the new snippet be $(\impmean, \impstd)$ and the
  AQP engine's raw answer and raw error to the new snippet be $(\estans_{n+1}, \beta_{n+1})$. Then,
  \[
    \impstd \le \beta_{n+1}
  \]
  and the equality occurs when the raw error is zero, or when \dbl's query synopsis is
  empty,
  or when \dbl's model-based answer is rejected by the model validation step.
  \label{thm:accuracy}
\end{theorem}
\endgroup

\begin{proof}
  Recall that $(\impmean, \impstd)$ is set either to \dbl's model-based answer/error, \ie, $(\mans,
  \mstd)$, or to the AQP system's raw answer/error, \ie, $(\estans_{n+1}, \beta_{n+1})$, depending
  on the result of the model validation. In the latter case, it is trivial that $\impstd \le
  \beta_{n+1}$, and hence it is enough to show that $\mstd \le \beta_{n+1}$.

  Computing $\mstd$ involves 
 an inversion of the covariance matrix $\Sigma_{n+1}$, where
  $\Sigma_{n+1}$ includes
  the $\beta_{n+1}$ term on one of its diagonal entries. We show $\mstd
  \le \beta_{n+1}$ by directly simplifying $\mstd$ into the form that involves $\beta_{n+1}$ and other
  terms.

  We first define notations. Let $\Sigma$ be the covariance matrix of the vector of random variables $(\rans_1,
  \ldots,$ $\rans_{n+1}, \rtans_{n+1})$;  $\kvec_n$ be a column vector of length
  $n$ whose $i$-th element is the $(i,n+1)$-th entry of $\Sigma$;  $\Sigma_n$
  be an $n \times n$ submatrix of $\Sigma$ that consists of $\Sigma$'s first $n$
  rows/columns; $\tqcov^2$ be a scalar value at the $(n+2,n+2)$-th entry of
  $\Sigma$; and $\estansvec_{n}$ be a column vector $(\rawans_1, \ldots, \rawans_{n})^\intercal$.

  Then, we can express $\kvec_{n+1}$ and $\Sigma_{n+1}$ in
  \cref{eq:infer,eq:impstd} in block forms as follows:
  \begin{align*}
    \kvec_{n+1} = \begin{pmatrix}
      \kvec_n \\ \tqcov^2
    \end{pmatrix}, \quad
    \Sigma_{n+1} = \begin{pmatrix}
      \Sigma_n  & \kvec_n \\
      \kvec_n^\intercal & \tqcov^2 + \beta_{n+1}^2
    \end{pmatrix}, \quad
    \estansvec_{n+1} = \begin{pmatrix}
      \estansvec_{n} \\ \rawans_{n+1}
    \end{pmatrix}
  \end{align*}
  Here, it is important to note that $\kvec_{n+1}$ can be expressed in terms of 
  $\kvec_n$ and $\tqcov^2$ because $(i,n+1)$-th element of $\Sigma$ and
  $(i,n+2)$-th element of $\Sigma$ have the same values for $i = 1, \ldots,
  n$. They have the same values because the covariance between $\rans_i$ and
  $\rans_{n+1}$ and the covariance between $\rans_i$ and $\rtans_{n+1}$ are same
  for $i = 1, \ldots, n$ due to \cref{eq:query_cov2}.

  Using the formula of block matrix inversion~\cite{mat_inverse},
  we can obtain the following alternative forms of \cref{eq:infer,eq:impstd} (here, we assume zero
  means to simplify the expressions):
  \begin{align}
    \gamma^2 &= \tqcov^2 - \kvec_n^\intercal \Sigma_{n}^{-1} \kvec_n, \qquad
    \theta = \kvec_n^\intercal \Sigma_{n}^{-1} \estansvec_{n}
    \label{eq:model_only} \\
    \mans &= 
    \frac{\beta_{n+1}^2 \cdot \theta + \gamma^2 \cdot \estans_{n+1}}{\beta_{n+1}^2 + \gamma^2},
    \qquad
    \mstd^2 = \frac{\beta_{n+1}^2 \cdot \gamma^2}{\beta_{n+1}^2 + \gamma^2}
    \label{eq:infer2} 
  \end{align}

  Note that $\mstd^2 < \beta_{n+1}$ for $\beta_{n+1} > 0$ and $\gamma^2 < \infty$, and $\mstd^2 =
  \beta_{n+1}$ if $\beta_{n+1} = 0$ or $\gamma^2 \rightarrow \infty$.
\end{proof}


\ignore{
Interestingly, the above theorem still holds even if the relation is subject to
	data additions (see \cref{sec:append}).

In proving the above theorem, we observe an interesting fact. In
\cref{eq:model_only,eq:infer2}, the raw answer and the error to the new snippet are not required for
computing the value of $\gamma^2$ and $\theta$. This observation helps to
keep \dbl's inference process at
query time fast, as described in the following lemma.
}

\begin{lemma}
  The time complexity of \dbl's inference is
  \rev{C2.1f}{$O(N^{max}\cdot l \cdot n^2 )$}
  \rev{C9}{The space complexity of \dbl is $O(n \cdot N^{max} + n^2)$,
    where $n \cdot N^{max}$ is the size of the query
    snippets and $n^2$ is the size of the precomputed covariance matrix.}
  \label{lemma:time}
\end{lemma}

\begin{proof}
  It is enough to prove that the computations of a model-based answer and a
  model-based error can be performed in $O(n^2)$ time, where $n$ is the number
  of past query snippets. Note that this is clear from
  \cref{eq:infer2,eq:model_only}, because the computation of $\Sigma_n^{-1}$
  involves only the past query snippets. For computing $\gamma^2$, multiplying
  $\kvec_n$, a precomputed $\Sigma_{n}^{-1}$, and $\kvec_n$ takes $O(n^2)$ time.
  Similarly for $\theta$ in \cref{eq:model_only}
\end{proof}

\rev{C2.1f,C9}{These results imply that the domain sizes of dimension attributes do not affect
\dbl's computational overhead.
This is because \dbl analytically computes the covariances between pairs of
query answers without individually computing inter-tuple covariances
(\cref{sec:data_stat}).}

\section{Verdict Process Summary}
\label{sec:summary}

\begin{algorithm}[t]
  \DontPrintSemicolon
  \KwIn{$Q_n$ including $(q_i, \estans_i, \beta_i)$ for $i = 1, \ldots, n$}
  \KwOut{$Q_n$ with new model parameters and precomputed matrices}

  \SetKwFunction{learn}{learn}
  \SetKwFunction{covans}{covariance}

  \BlankLine
  \ForEach{aggregate function $g$ in $Q_n$}{
    $(l_{g,1}, \ldots, l_{g,l}, \sigma_g^2)$ $\leftarrow$ \learn{$Q_n$}
    \tcp*{\cref{sec:learn}}

    \BlankLine
    \tcp{$\Sigma_{(i,j)}$ indicates $(i,j)$-element of $\Sigma$}
    \For{$(i,j) \leftarrow (1, \ldots, n) \times (1, \ldots, n)$}{
      \tcp{\cref{eq:query_cov2}}
      $\Sigma_{(i,j)} \leftarrow$ \covans{$q_i, q_j; \, l_{g,1}, \ldots, l_{g,l}, \sigma_g^2$}
    }

    Insert $\Sigma$ and $\Sigma^{-1}$ into $Q_n$ for $g$
  }

  \Return $Q_n$

  \caption{\dbl offline process}
  \label{algo:offline}
\end{algorithm}

\begin{algorithm}[t]
  \DontPrintSemicolon
  \KwIn{New query snippets $q_{n+1}, \ldots, q_{n+b}$,\\
  \qquad \quad Query synopsis $Q_n$
  }
  \KwOut{$b$ number of improved answers and improved errors \\
  \qquad \qquad \qquad $\{(\widehat{\theta}_{n+1}, \widehat{\beta}_{n+1}), \ldots,
    (\widehat{\theta}_{n+b}, \widehat{\beta}_{n+b})\}$, \\
  \qquad \qquad Updated query synopsis $Q_{n+b}$
  }

  \SetKwData{fcount}{fc}
  \SetKwFunction{safeguard}{valid}
  \SetKwFunction{aqp}{AQP}
  \SetKwFunction{improve}{inference}
  \SetKwFunction{supp}{supported}
  \BlankLine

  $\fcount \leftarrow$ number of distinct aggregate functions in new queries

  \BlankLine
  \tcc{The new query (without decomposition) is sent to the AQP engine in
  practice.}
  $\{(\estans_{n+1}, \beta_{n+1}), \ldots, (\estans_{n+b}, \beta_{n+b})\}$
  $\leftarrow$ \aqp{$q_{n+1}, \ldots, q_{n+b}$}


  \BlankLine
  \tcp{improve up to $N^{max}$ rows}
  \For{$i \leftarrow 1, \ldots, (\fcount \cdot N^{max})$}{

    \BlankLine
    \tcp{model-based answer/error}
    \tcp{(\cref{eq:infer,eq:impstd})}
    $(\ddot{\theta}_{n+i}, \ddot{\beta}_{n+i}) \leftarrow$
    \improve{$\estans_{n+i}, \beta_{n+i}, Q_n$}
    \BlankLine

    \tcp{model validation (\cref{sec:safeguard})}
    $(\widehat{\theta}_{n+i}, \widehat{\beta}_{n+i}) \leftarrow$
    \leIf{\safeguard{$\ddot{\theta}_{n+i}$, $\ddot{\beta}_{n+i}$}}
    {
      $(\ddot{\theta}_{n+i}, \ddot{\beta}_{n+i})$
    }{
      $(\estans_{n+i}, \beta_{n+i})$
    }

    Insert $(q_{n+i}, \estans_{n+i}, \beta_{n+i})$ into $Q_{n}$
  }

  \BlankLine
  \For{$i \leftarrow (\fcount \cdot N^{max} + 1), \ldots, b$}{
      $(\widehat{\theta}_{n+i}, \widehat{\beta}_{n+i}) \leftarrow (\estans_{n+i}, \beta_{n+i})$
  }

  \tcp{\dbl overhead ends}

  \Return $\{(\widehat{\theta}_{n+1}, \widehat{\beta}_{n+1}), \ldots,
    (\widehat{\theta}_{n+b}, \widehat{\beta}_{n+b})\}, Q_n$

  \caption{\dbl runtime process}
  \label{algo:online}
\end{algorithm}

\rev{C8}{
In this section, we summarize \dbl's offline and online processes.
Suppose the query synopsis $Q_n$
contains a total of $n$ query snippets from past query processing,
and a new query is decomposed into $b$ query
snippets; we denote the new query snippets in the new query by $q_{n+1}, \ldots,
q_{n+b}$.

\ph{Offline processing} 
\Cref{algo:offline} summarizes \dbl's offline process. It consists of learning
correlation parameters  and computing covariances between all pairs of
past query snippets.

\ph{Online processing}
\Cref{algo:online} summarizes \dbl's runtime process. Here, we
assume the new query is a supported query; otherwise, \dbl simply forwards the
AQP engine's query answer to the user.


}


\section{Deployment Scenarios}
\label{sec:overview:deployment}



\dbl is designed to support a large class of AQP engines.
However, depending on the type
of AQP engine used, \dbl may provide both speedup and error reduction, or only error
reduction.

\begin{enumerate}[nolistsep,itemsep=3pt,wide,labelwidth=!,labelindent=0pt,topsep=3pt,partopsep=3pt,partopsep=3pt]
  \item  \textbf{AQP engines that support online aggregation}
    \cite{online-agg,g-ola,online-agg-mr1,zeng2016iolap}:
    Online aggregation continuously refines its approximate answer as new tuples
    are processed, until users are satisfied with the current accuracy or when
    the entire dataset is processed.  In these types of engines, every time the
    online aggregation provides an updated answer (and error estimate), \dbl
    generates an improved answer with a higher accuracy (by paying small
    runtime overhead).      As soon as this
    accuracy meets the user requirement, the online aggregation can be stopped.
    With \dbl, the online aggregation's continuous processing will stop
    earlier than it would without \dbl.
    This is because
     \dbl reaches a target error bound much earlier
    by combining its model with the raw answer of the AQP engine.


  \item \textbf{AQP engines that support time-bounds}  
    \cite{mozafari_eurosys2013,surajit-optimized-stratified,aggr_oracle,dbo,sample_subset,mozafari_sigmod2014_abm,easy_bound_bootstrap}:
    Instead of continuously refining approximate answers and  reporting them to the user, 
    	these engines simply take a time-bound from the user,
      and then they predict the largest sample size that they can
      process within the requested  time-bound; 
      thus, they minimize error bounds within the allotted time.
    For these engines, \dbl simply replaces the user's original time
    bound $t_1$ with a slightly smaller value $t_1-\epsilon$
    before passing it down to the AQP
    engine, where $\epsilon$ is the time needed by \dbl for inferring the improved answer and improved error.
    Thanks to the efficiency of \dbl's inference, $\epsilon$ is typically a small value,
    \eg, a few milliseconds (see \cref{sec:exp:overhead}).
Since \dbl's inference brings larger accuracy improvements on average compared to the
  benefit of processing more tuples within the $\epsilon$ time,
    \dbl achieves significant error reductions over traditional AQP engines.

\end{enumerate}

In this paper, we use an online aggregation engine to demonstrate \dbl's  both
speedup and error reduction capabilities (\cref{sec:exp}).
However, for interested readers, we also provide evaluations on a time-bound
engine~\cref{sec:time-bound}.

Some AQP engines also support error-bound queries but do not 
	offer an online aggregation interface~\cite{mozafari_pvldb2012,mozafari_sigmod2016_demo,mozafari_cidr2017}. 
	For these engine, 
	\dbl currently only benefits their time-bound queries,
		leaving their answer to error-bound queries  unchanged.
Supporting the latter would require either adding an online aggregation interface to the AQP engine,
	or a tighter integration of \dbl and the AQP engine itself.	
Such modifications are beyond the scope of this paper, as one of our design goals is to treat
	the underlying AQP engine as a black box (\cref{fig:overview:workflow}),
		so that \dbl can be used alongside a larger number of existing engines.

\rev{C3}{
Note that \dbl's inference mechanism is not affected by the specific AQP
engine used underneath, as long as the conditions in \cref{sec:infer} hold, namely the error estimate $\beta^2$
is the expectation of the squared deviation of the approximate answer from the
exact answer.
However, the AQP engine's runtime overhead (\eg, query parsing and planning) may affect
\dbl's overall benefit in relative terms. For example, if the query parsing
amount to 90\% of the overall query processing time, even if Verdict completely
eliminates the need for processing any data, the relative speedup will only be
1.0/0.9 = 1.11$\times$. However, Verdict is designed for data-intensive scenarios where
disk or network I/O is a sizable portion of the overall query processing time.}




\section{Experiments}
\label{sec:exp}

We conducted experiments to 
(i) quantify the  percentage of real-world queries
that  benefit from \dbl (\cref{sec:exp:query_analysis}),
(ii) study \dbl's average speedup and error reductions over an AQP engine (\cref{sec:exp:speedup}),
(iii) test the reliability of \dbl's error bounds
(\cref{sec:exp:stat_error}),
(iv) measure \dbl's computational overhead and memory footprint
(\cref{sec:exp:overhead}), and
(v) study the impact of different workloads and data distributions on \dbl's
effectiveness  (\cref{sec:exp:parameter}).
In summary, our results indicated the following:
\begin{itemize}[leftmargin=5mm,itemsep=1pt,nolistsep,topsep=0pt,partopsep=3pt,partopsep=3pt]
  \item \dbl supported a large fraction (73.7\%) of aggregate queries in a real-world workload, and
    produced significant speedups (up to 23.0$\times$) compared to a sample-based AQP solution.
  \item Given the same processing time, \dbl reduces the baseline's approximation error on average
    by 75.8\%--90.2\%. 
  \item \dbl's runtime overhead was $<$10 milliseconds on average (0.02\%--0.48\% of total
    time) and its memory footprint was negligible.
    \item \dbl's approach was robust against various workloads and data distributions.
\end{itemize}

We also have supplementary experiments in \cref{sec:more:exp}.
\cref{sec:exp:contribution} shows the benefits of model-based inference in
comparison to a strawman approach, which simply caches all past query answers.
\cref{sec:time-bound} demonstrates \dbl's benefit for time-bound
AQP engines.




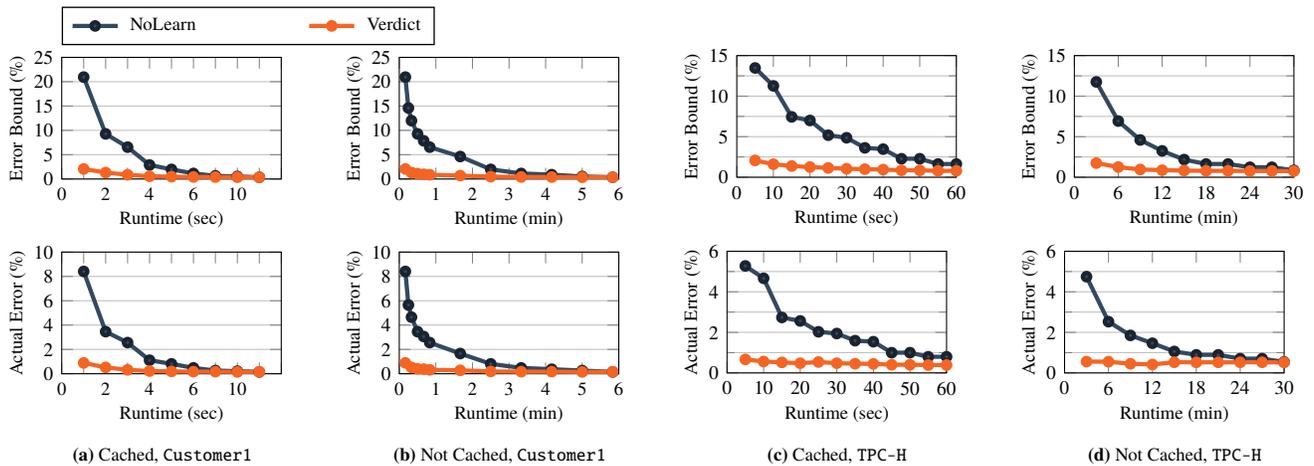
\begin{figure*}

  \pgfplotsset{laterr/.style={
      width=45mm,
      height=32mm,
      ylabel near ticks,
      ylabel shift=-3pt,
      xlabel shift=-2pt,
      xlabel near ticks,
      ylabel style={align=center},
      xlabel=Runtime (sec),
      ylabel=Error Bound (\%),
      legend style={
        at={(0,1.1)},anchor=south west,column sep=2pt,
        draw=black,fill=none,font=\scriptsize,line width=.5pt,
        /tikz/every even column/.append style={column sep=40pt}
      },
      legend columns=2,
      every axis/.append style={font=\scriptsize},
      ymajorgrids,
      minor grid style=lightgray,
      yminorgrids,
  }}

  \pgfplotsset{graydot/.style={
      draw=darkteal,
      mark=*,
      mark options={scale=0.7,draw=black!50!darkteal,fill=darkteal},
      ultra thick,
  }}
  \pgfplotsset{reddot/.style={
      draw=darkorange,
      mark=*,
      mark options={scale=0.7,draw=darkorange,fill=darkorange},
      ultra thick,
  }}

  \centering
  \begin{subfigure}[b]{0.24\textwidth}
    \begin{tikzpicture}
      \begin{axis}[
          laterr,
          xmin=0,
          xmax=10,
          xtick={0, 1, ..., 10},
          xticklabels={0, , 2, , 4, , 6, , 10},
          ymin=0,
          ymax=25,
          ytick={0, 5, ..., 25},
        ]
        \addplot[graydot]
        table[x=x,y=y]{
          x	y
          1	20.95
          2	9.27
          3	6.55
          4	2.88
          5	1.98
          6	1.14
          7	0.65
          8	0.52
          9	0.36
        };

        \addplot[reddot]
        table[x=x,y=y]{
          x	y
          1	2.06
          2	1.32
          3	0.87
          4	0.56
          5	0.48
          6	0.42
          7	0.40
          8	0.37
          9	0.36
        };

        \addlegendentry{\nol};
        \addlegendentry{\dbl};

      \end{axis}
    \end{tikzpicture}

    \begin{tikzpicture}
      \begin{axis}[
          laterr,
          xmin=0,
          xmax=10,
          xtick={0, 1, ..., 10},
          xticklabels={0, , 2, , 4, , 6, , 10},
          ymin=0,
          ymax=10,
          ytick={0, 2, ..., 10},
          ylabel=Actual Error (\%),
        ]

        \addplot[graydot]
        table[x=x,y=y]{
          x	y
          1	8.41
          2	3.46
          3	2.56
          4	1.11
          5	0.81
          6	0.48
          7	0.26
          8	0.21
          9	0.15
        };

        \addplot[reddot]
        table[x=x,y=y]{
          x	y
          1	0.89
          2	0.52
          3	0.32
          4	0.22
          5	0.19
          6	0.17
          7	0.17
          8	0.15
          9	0.15
        };


      \end{axis}
    \end{tikzpicture}
    \caption{Cached, \vertica}
  \end{subfigure}
  ~
  \begin{subfigure}[b]{0.24\textwidth}
    \begin{tikzpicture}
      \begin{axis}[
          laterr,
          xmin=0,
          xmax=360,
          xtick={0, 60, ..., 360},
          xticklabels={0, 1, 2, 3, 4, 5, 6},
          ymin=0,
          ymax=25,
          ytick={0, 5, ..., 25},
          xlabel=Runtime (min),
        ]
        \addplot[graydot]
        table[x=x,y=y]{
          x	y
          10	20.95
          15	14.60
          20	11.99
          30	9.27
          40	7.86
          50	6.55
          100	4.62
          150	1.98
          200	1.14
          250	0.89
          300	0.52
          350	0.36
        };

        \addplot[reddot]
        table[x=x,y=y]{
          x	y
          10	2.06
          20	1.32
          30	1.10
          40	0.97
          50	0.87
          100	0.70
          150	0.48
          200	0.42
          250	0.40
          300	0.37
          350	0.36
        };


      \end{axis}
    \end{tikzpicture}

    \begin{tikzpicture}
      \begin{axis}[
          laterr,
          xmin=0,
          xmax=360,
          xtick={0, 60, ..., 360},
          xticklabels={0, 1, 2, 3, 4, 5, 6},
          ymin=0,
          ymax=10,
          ytick={0, 2, ..., 10},
          ylabel=Actual Error (\%),
          xlabel=Runtime (min),
        ]

        \addplot[graydot]
        table[x=x,y=y]{
          x	y
          10	8.41
          15	5.65
          20	4.65
          30	3.46
          40	3.06
          50	2.56
          100	1.66
          150	0.81
          200	0.48
          250	0.37
          300	0.26
          350	0.15
        };

        \addplot[reddot]
        table[x=x,y=y]{
          x	y
          10	0.89
          20	0.52
          30	0.41
          40	0.38
          50	0.32
          100	0.28
          150	0.19
          200	0.17
          250	0.17
          300	0.15
          350	0.15
        };


      \end{axis}
    \end{tikzpicture}
    \caption{Not Cached, \vertica}
  \end{subfigure}
  ~
  \begin{subfigure}[b]{0.24\textwidth}    
    \begin{tikzpicture}
      \begin{axis}[
          laterr,
          xmin=0,
          xmax=60,
          xtick={0, 10, ..., 60},
          ymin=0,
          ymax=15,
          ytick={0, 5, 10, 15},
          minor y tick num={1},
        ]
        \addplot[graydot]
        table[x=x,y=y]{
          x	y
          5	13.47316
          10	11.26536
          15	7.4426
          20	6.99902
          25	5.18082
          30	4.86654
          35	3.62274
          40	3.4636
          45	2.2767
          50	2.2767
          55	1.63136
          60	1.63136
        };

        \addplot[reddot]
        table[x=x,y=y]{
          x	y
          5	2.06918
          10	1.60634
          15	1.3799
          20	1.25266
          25	1.14332
          30	1.04378
          35	0.99464
          40	0.9429
          45	0.85196
          50	0.85196
          55	0.7882
          60	0.7882
        };


      \end{axis}
    \end{tikzpicture}

    \begin{tikzpicture}
      \begin{axis}[
          laterr,
          xmin=0,
          xmax=60,
          xtick={0, 10, ..., 60},
          ymin=0,
          ymax=6,
          ytick={0, 2, 4, 6},
          minor y tick num={1},
          ylabel=Actual Error (\%),
        ]
        \addplot[graydot]
        table[x=x,y=y]{
          x	y
          5	5.27514
          10	4.6661
          15	2.73418
          20	2.56378
          25	2.02726
          30	1.94226
          35	1.59076
          40	1.53864
          45	0.99782
          50	0.99782
          55	0.79562
          60	0.79562
        };

        \addplot[reddot]
        table[x=x,y=y]{
          x	y
          5	0.66526
          10	0.56224
          15	0.51804
          20	0.47798
          25	0.53796
          30	0.47874
          35	0.45348
          40	0.44482
          45	0.40374
          50	0.40374
          55	0.38742
          60	0.38742
        };


      \end{axis}
    \end{tikzpicture}

    \caption{Cached, \tpch}
  \end{subfigure}
  ~
  \begin{subfigure}[b]{0.24\textwidth}
    \begin{tikzpicture}
      \begin{axis}[
          laterr,
          xmin=0,
          xmax=30,
          xtick={0, 6, ..., 30},
          ymin=0,
          ymax=15,
          ytick={0, 5, 10, 15},
          minor y tick num={1},
          xlabel=Runtime (min),
        ]
        \addplot[graydot]
        table[x=x,y=y]{
          x	y
          3	11.75456
          6	6.92426
          9	4.59428
          12	3.24928
          15	2.1755
          18	1.63136
          21	1.63136
          24	1.23058
          27	1.23058
          30	0.84648
        };

        \addplot[reddot]
        table[x=x,y=y]{
          x	y
          3	1.74378
          6	1.25266
          9	0.95332
          12	0.86916
          15	0.81694
          18	0.7882
          21	0.7882
          24	0.76982
          27	0.76982
          30	0.75252
        };


      \end{axis}
    \end{tikzpicture}

    \begin{tikzpicture}
      \begin{axis}[
          laterr,
          xmin=0,
          xmax=30,
          xtick={0, 6, ..., 30},
          ymin=0,
          ymax=6,
          ytick={0, 2, 4, 6},
          minor y tick num={1},
          xlabel=Runtime (min),
          ylabel=Actual Error (\%),
        ]
        \addplot[graydot]
        table[x=x,y=y]{
          x	y
          3	4.74774
          6	2.52744
          9	1.85026
          12	1.46354
          15	1.05742
          18	0.89044
          21	0.89044
          24	0.70342
          27	0.70342
          30	0.54038
        };

        \addplot[reddot]
        table[x=x,y=y]{
          x	y
          3	0.5579
          6	0.5534
          9	0.45142
          12	0.41416
          15	0.52608
          18	0.52176
          21	0.52176
          24	0.52414
          27	0.52414
          30	0.518
        };


      \end{axis}
    \end{tikzpicture}
    \caption{Not Cached, \tpch}
  \end{subfigure}

  \caption{The relationship (i) between runtime and error bounds (top row), and (ii)
    between runtime and actual errors (bottom row), for both systems: \nol and \dbl.}
  \label{fig:exp:laterr}
\end{figure*}

\subsection{Experimental Setup}
\label{sec:exp:setup}

\ph{Datasets and Query Workloads}
For our experiments, we used the three datasets described below:
\begin{enumerate}[leftmargin=5mm,itemsep=1pt,nolistsep]
  \item \vertica: This is a real-world query trace from
      one of the largest customers (anonymized) of a leading vendor of analytic
      DBMS.
    This dataset contains 310 tables and 15.5K
    timestamped queries issued between March 2011 and April 2012, 3.3K of which
    are analytical queries supported by Spark SQL. 
    We did not have the customer's original dataset but had access to their data
    distribution, which we used to generate a 536 GB dataset.

  \item \tpch: This is a well-known  analytical benchmark with 22 query types,
    21 of which  contain at least one aggregate
    function (including 2 queries with \texttt{min} or \texttt{max}).
    We used a scale factor of 100, i.e., the total data size was 100 GB. 
    We generated a total of 500 queries using \tpch's workload generator with
    its default settings.
    The queries in this dataset include joins of up to 6 tables.

  \item \synthetic: For more controlled experiments, we also generated large-scale synthetic
    datasets with different distributions (see \cref{sec:exp:parameter} for details).
\end{enumerate}

\ph{Implementation}
For comparative analysis, we implemented two systems on top of
Spark SQL~\cite{armbrust2015spark} (version 1.5.1):
\begin{enumerate}[leftmargin=5mm,itemsep=1pt,nolistsep]
  \item \nol: This system is an online aggregation engine that creates
   random samples of the original tables offline and splits
    them into multiple batches  of tuples. 
    To compute increasingly accurate
    answers to a new query, \nol first
    computes an approximate answer and its associated error bound on the first
    batch
    of tuples, 
    and then continues to refine its answer and error bound as
    it processes additional batches.
    \rev{C4}{\nol estimates its errors and computes confidence intervals using
    closed-forms (based on the central limit theorem). Error estimation based on
    the central limit
    theorem has been one of the
    most popular approaches in online aggregation
    systems~\cite{online-agg,g-ola,zeng2016iolap,kandula2016quickr} and other AQP
    engines~\cite{mozafari_eurosys2013,surajit-optimized-stratified,join_synopses}.}

\item \dbl: This system is an implementation of our proposed approach, which uses \nol as its AQP
    engine. In other words, each time \nol yields a raw answer and error, \dbl computes an
    improved answer and error using our proposed approach. 
    Naturally, \rev{C4}{\dbl incurs a (negligible) 
     runtime overhead,
    due to supported query check, query decomposition, and computation
    of improved answers;}
    	however, \dbl yields answers that are much more accurate in  general.
%
%
\ignore{	 when users append a
    special prefix (\texttt{l}) in front of regular aggregate functions. For
    instance, if an aggregate function \texttt{LSUM(X)} is specified inside a
    regular SQL statement in place of \texttt{SUM(X)}, \dbl inserts an extra
    query analysis step before running the query and modifies \nol's result
    set right before returning the query's answer to users.}
\end{enumerate}

\ph{Experimental Environment} We used a Spark cluster (for both \nol and \dbl) using 5 Amazon EC2 \texttt{m4.2xlarge} instances, 
each with 2.4 GHz Intel Xeon E5 processors (8 cores) and 32GB of memory. 
Our cluster also included SSD-backed
HDFS~\cite{shvachko2010hadoop} for Spark's data loading.  For experiments with cached datasets, 	
	 we distributed Spark's RDDs evenly across the nodes using Spark SQL DataFrame \texttt{repartition} function. 
	 \ignore{For \tpch, samples were
prejoined with other tables to reduce online query times. Note that samples were
significantly smaller than original tables, so the sizes of the joined tables
were also very small compared to original tables.}

\begin{table}[t]
  \centering
  \small

  \begin{tabular}{ l r r r }
    \hline
    \multirow{2}{*}{\textbf{Dataset}}  & \textbf{Total \# of Queries} &
    \textbf{\# of Supported} & \multirow{2}{*}{\textbf{Percentage}} \\
             & \textbf{with Aggregates} &
    \textbf{Queries} & \\
    \Xhline{2\arrayrulewidth}
    \vertica & 3,342 & 2,463 & 73.7\% \\ \hline
    \tpch    & 21    & 14    & 63.6\% \\ \hline
  \end{tabular}

  \caption{Generality of \dbl. \dbl supports a large fraction of real-world
  and benchmark queries.}

  \label{tab:exp:generality}
  \vspace{\neggap}
\end{table}

\vspace{5mm}
\subsection{Generality of \secdbl}
\label{sec:exp:query_analysis}

To quantify the generality of our approach, we measured the coverage of 
	 our supported queries in practice.
We
analyzed the real-world SQL queries in \vertica. 
From  the original 15.5K queries, Spark SQL was able to
	 process 3.3K of the aggregate queries. 
Among those 3.3K queries, \dbl supported 2.4K queries, i.e., 73.7\% of the analytical queries could benefit from \dbl.
In addition, we analyzed the 21 \tpch queries and found 14 queries supported by \dbl. 
Others could not be supported due to textual filters or disjunctions in the \texttt{where} clause.
These statistics are summarized in
Table~\ref{tab:exp:generality}.
This analysis proves that \dbl can support a large class of analytical queries in practice. 
Next, we quantified the extent to which these supported queries
benefitted from Verdict.

\subsection{Speedup and Error Reduction}
\label{sec:exp:speedup}

In this section, we first study the relationship between the processing time and
the size of error bounds for both systems, i.e., \nol and \dbl. Based
on this study, we then
analyze \dbl's  speedup and error reductions over \nol.

In this experiment, we used each of \vertica and \tpch datasets in two
different settings. In one setting,  all samples were cached in the memories of
the cluster, while in the second, the data had to be read from SSD-backed
HDFS.

We allowed both systems to process half of the queries (since \vertica queries
were timestamped, we used the first half).  While processing those queries, \nol
simply returned the query answers, but \dbl also kept the queries and their
answers in its query synopsis.
\rev{C5}{After processing those queries, \dbl
(i) precomputed the matrix inversions and (ii)
learned the correlation parameters. 
 The
matrix inversions took 1.6 seconds in total; the correlation parameter
learning took 23.7 seconds for \tpch and 8.04 seconds for \vertica. The learning
process was
relatively faster for \vertica since most of the queries included
\texttt{COUNT(*)} for which each attribute did not require a separate learning.
This offline training time for both workloads was
comparable to the time needed for
running only a single approximate query (\cref{tab:exp:speedup}).}

For the second half of the queries, we recorded both systems' query
processing times (\ie, runtime), approximate query answers, and
error bounds. Since both \nol and \dbl   are online aggregation systems, and
\dbl produces improved answers for every answer returned from \nol, both systems
naturally
produced more accurate answers (\ie,  answers with smaller error bounds) as
query processing continued.  Approximate query engines, including both \nol and
\dbl, are only capable of producing expected errors in terms of error bounds.
However, for analysis, we also computed the actual errors by comparing those
approximate answers against the exact answers.  
In the following, we report their relative errors.

\Cref{fig:exp:laterr} shows the relationship between runtime and  average error
bound (top row) and the relationship between runtime and 
average actual
error (bottom row). Here, we also considered two cases: when the entire data is cached in memory 
and when it resides on SSD.
In all experiments, the runtime-error graphs exhibited a consistent pattern: (i) \dbl
produced smaller errors even when runtime was very large, and
(ii) \dbl showed faster runtime for the same target errors.
Due to the asymptotic nature of errors, achieving extremely accurate
answers (\eg, less than 0.5\%) required relatively long processing time even for
\dbl.

\begin{table}[t]
  \centering
  \small

  \renewcommand{\arraystretch}{1.2}

  \begin{tabular}{c l r r r r r }
    \hline
    & \multirow{2}{*}{\textbf{Cached?}} & \textbf{Error}
    & \multicolumn{3}{c}{\textbf{Time Taken}}  & \multirow{2}{*}{\textbf{Speedup}} \\ 
                                    & & \textbf{Bound}  & & \nol     & \dbl       & \\ \Xhline{2\arrayrulewidth}
    \parbox[t]{2mm}{\multirow{4}{*}{\rotatebox[origin=c]{90}{\vertica}}}
    & \multirow{2}{*}{Yes} & 2.5\%  & & 4.34 sec  & 0.57 sec & \textbf{7.7$\times$} \\
    &                      & 1.0\%  & & 6.02 sec  & 2.45 sec & \textbf{2.5$\times$} \\ \cline{2-7}
    & \multirow{2}{*}{No}  & 2.5\%  & & 140 sec   & 6.1 sec  & \textbf{23.0$\times$} \\
    &                      & 1.0\%  & & 211 sec   & 37 sec   & \textbf{5.7$\times$} \\ \hline
    \parbox[t]{2mm}{\multirow{4}{*}{\rotatebox[origin=c]{90}{\tpch}}}
    & \multirow{2}{*}{Yes} & 4.0\%  & & 26.7 sec  & 2.9 sec  & \textbf{9.3$\times$} \\
    &                      & 2.0\%  & & 34.2 sec  & 12.9 sec & \textbf{2.7$\times$} \\ \cline{2-7}
    & \multirow{2}{*}{No}  & 4.0\%  & & 456 sec   & 72 sec   & \textbf{6.3$\times$} \\
    &                      & 2.0\%  & & 524 sec   & 265 sec  & \textbf{2.1$\times$} \\ \hline \hline
    & \multirow{2}{*}{\textbf{Cached?}} & \multirow{2}{*}{\textbf{Runtime}} 
    & \multicolumn{3}{c}{\textbf{Achieved Error Bound}}  & \textbf{Error} \\ 
    &  &         & & \nol     & \dbl              & \textbf{Reduction}  \\ \Xhline{2\arrayrulewidth}
    \parbox[t]{2mm}{\multirow{4}{*}{\rotatebox[origin=c]{90}{\vertica}}}
    & \multirow{2}{*}{Yes} & 1.0 sec & & 21.0\%  & 2.06\%  & \textbf{90.2\%} \\
    &                      & 5.0 sec & & 1.98\%  & 0.48\%  & \textbf{75.8\%} \\ \cline{2-7}
    & \multirow{2}{*}{No}  & 10 sec  & & 21.0\%  & 2.06\%  & \textbf{90.2\%} \\
    &                      & 60 sec  & & 6.55\%  & 0.87\%  & \textbf{86.7\%} \\ \hline
    \parbox[t]{2mm}{\multirow{4}{*}{\rotatebox[origin=c]{90}{\tpch}}}
    & \multirow{2}{*}{Yes} & 5.0 sec & & 13.5\%  & 2.13\%  & \textbf{84.2\%} \\
    &                      & 30 sec  & & 4.87\%  & 1.04\%  & \textbf{78.6\%} \\ \cline{2-7}
    & \multirow{2}{*}{No}  & 3.0 min & & 11.8\%  & 1.74\%  & \textbf{85.2\%} \\
    &                      & 10 min  & & 4.49\%  & 0.92\%  & \textbf{79.6\%} \\ \hline
  \end{tabular}

  \caption{Speedup and error reductions by \dbl compared to \nol.}
  \label{tab:exp:speedup}
  \vspace{\neggap}
\end{table}

Using these results, we also analyzed \dbl's speedups and error
reduction over \nol. For speedup, we compared how long each system took
until it reached a target error bound. 
For
error reduction, we compared the lowest error bounds that each system
 produced within a fixed allotted time. \Cref{tab:exp:speedup}
reports the results for each combination of dataset and location (in memory or on SSD).

For the \vertica dataset, \dbl achieved a larger speedup when the data was
stored on
SSD (up to 23.0$\times$) compared to when it was fully cached in
memory
(7.7$\times$). The reason was that, for cached data,
the I/O time was no longer the dominant factor and 
 Spark SQL's default overhead (e.g., parsing the query and reading the catalog) 
  accounted for a considerable portion of the total data processing time.	
For  \tpch , on the
contrary, the speedups were smaller when the data was stored on SSD. 
This difference stems from the different query forms between
\vertica and \tpch. The \tpch dataset includes queries that join
	 several tables, some of which are large tables
   that  were not sampled by \nol. (Similar to most sample-based AQP engines, \nol only samples fact tables, not dimension
tables.)
Consequently, those large tables had to be read each time \nol
processed such a query. When the data resided on SSD, loading those tables became
	a major bottleneck that could not be reduced by \dbl (since they were not sampled).
However, on average, \dbl still achieved an impressive 6.3$\times$ speedup over \nol. 
In general, \dbl's speedups over \nol
reduced as the target error bounds became smaller; however, even for 1\%
target error bounds, \dbl achieved an average of up to 5.7$\times$ speedup over \nol.

\cref{tab:exp:speedup} also reports \dbl's error reductions over \nol.
For all target runtime budgets we examined, \dbl achieved massive error reductions
compared to \nol.

The performance benefits of \dbl depends on several important factors, such as the
accuracy of past query answers and workload characteristics. These factors are further
studied in \cref{sec:exp:parameter,sec:exp:contribution}.

\subsection{Confidence Interval Guarantees}
\label{sec:exp:stat_error}

To confirm the validity of \dbl's error bounds, we configured \dbl to produce error bounds at 95\%
confidence and compared them to the actual errors.
We ran \dbl
  for an amount of time long enough to sufficiently collect error bounds of
various sizes.

\begin{figure}[t]
  \pgfplotsset{statcorrect/.style={
      width=70mm,
      height=32mm,
      xmin=0.8,
      xmax=6.2,
      ymin=0,
      ymax=30,
      xtick={1, 2, 3, 4, 5, 6},
      xticklabels={1\%, 2\%, 4\%, 8\%, 16\%, 32\%},
      ytick={0, 5, ..., 30},
      ylabel near ticks,
      ylabel shift=-2pt,
      xlabel shift=-2pt,
      xlabel near ticks,
      ylabel style={align=center},
      xlabel=\dbl Error Bound,
      ylabel=Actual Error (\%),
      legend style={
        at={(0,1)},anchor=south west,column sep=3pt,
        draw=none,font=\scriptsize,fill=none,line width=.5pt},
      legend columns=1,
      every axis/.append style={font=\scriptsize},
      tick style={draw=none},
      ymajorgrids,
      xmajorgrids,
      legend cell align=left,
  }}

  \pgfplotsset{graybar/.style={
      fill=none,
      draw=darkteal,
      ultra thick,
      mark=none,
  }}

  \tikzset{errmark/.style={
      draw=darkorange,
      ultra thick,
  }}

  \centering
  \begin{tikzpicture}
    \begin{axis}[
        statcorrect,
      ]



      \addplot[graybar,
          error bars/.cd,
          y dir=both,
          y explicit,
          error bar style={line width=1pt,darkteal},
          error mark options={
              rotate=90,
              darkteal,
              mark size=4pt,
              line width=1pt
          },
      ]
      coordinates {
        (1, 0.099) +-  (0.253, 0.095)
        (2, 0.899) +-  (2.264, 0.874)
        (3, 1.054) +-  (1.376, 0.878)
        (4, 3.490) +-  (5.338, 3.176)
        (5, 6.027) +-  (8.906, 5.506)
        (6, 11.144) +- (8.008, 10.372)
      };




    \end{axis}
  \end{tikzpicture}

  \caption{The comparison between \dbl's error bound at 95\% confidence and
    the actual error distribution (5th, 50th, and 95th percentiles are reported for actual error
    distributions).}
  \label{fig:exp:staterror}
  \vspace{\neggap}
\end{figure}
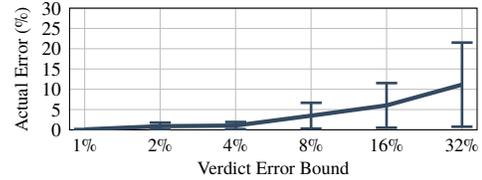

By definition, the error bounds at 95\% confidence are probabilistically correct if the actual errors are smaller than
the error bounds in at least 95\% of the cases.
Figure~\ref{fig:exp:staterror} shows the 5th percentile, median, and 95th
percentile of the actual errors for different sizes of error bounds (from 1\% to 32\%).
 In all cases, the 95th
percentile of the actual errors were lower than the error bounds produced by \dbl,
	which confirms the probabilistic correctness of \dbl's error bound guarantees.

\vspace{5mm}
\subsection{Memory and Computational Overhead}
\label{sec:exp:overhead}

In this section, we study \dbl's additional memory footprint (due to query synopsis) 
and its runtime overhead (due to inference).
\rev{C6}{
The total memory footprint of the query synopsis was 5.79MB for \tpch and 18.5MB
for \vertica workload (23.2KB per-query for \tpch and 15.8KB per-query for
\vertica).
This included past
queries in parsed forms, model parameters,
covariance matrices, and the inverses of those
covariance matrices.}
The size of query synopsis was small because \dbl does not retain any of the input tuples.


To measure \dbl's runtime overhead, we recorded  the time spent for its regular
query processing (\ie, \nol) and the  additional time spent
 for the inference and updating the final answer.
 As summarized in Table~\ref{tab:exp:overhead},  the
  runtime overhead of \dbl was negligible
compared to the overall query processing time.
This is because  multiplying a vector by a $C_g\times C_g$ matrix 
does not take much time compared to regular query planning, processing, and
network commutations among the distributed nodes.
(Note that $C_g$$=$$2,000$ by
default; see \cref{sec:overview:internal}.)

\begin{table}[!t]
  \centering
  \small

  \vspace{2mm}

  \begin{tabular}{|l|r r|r r|}
    \hline
    \textbf{Latency }    & \multicolumn{2}{c|}{\textbf{Cached}}
    & \multicolumn{2}{c|}{\textbf{Not Cached}}     \\ \Xhline{2\arrayrulewidth}
    \nol                 & 2.083 sec &          & 52.50 sec &  \\ \hline
    \dbl                 & 2.093 sec &          & 52.51 sec &  \\ \hline
    \textbf{Overhead}    & 0.010 sec  & (0.48\%) & 0.010 sec & (0.02\%) \\ \hline
  \end{tabular}

  \vspace{-1mm}
  \caption{The runtime overhead of \dbl.}
  \label{tab:exp:overhead}

  \vspace{\neggap}
\end{table}

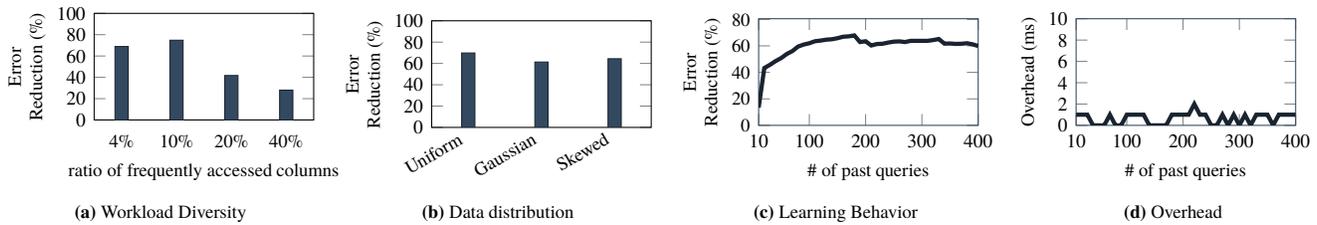
\begin{figure*}

  \pgfplotsset{param1/.style={
      width=45mm,
      height=30mm,
      ymin=0,
      ymax=100,
      bar width=5pt,
      ylabel near ticks,
      ylabel shift=-2pt,
      xlabel near ticks,
      ylabel style={align=center},
      ylabel=Error\\ Reduction (\%),
      legend style={
        at={(0,1.2)},anchor=south west,column sep=3pt,
        draw=none,font=\scriptsize,fill=none,font=\tiny,line width=.5pt},
      legend columns=2,
      every axis/.append style={font=\scriptsize},
      ybar,
      area legend,
      xtick style={draw=none},
      x tick label style={yshift=2pt},
  }}

  \pgfplotsset{parambar/.style={
      draw=black!50!darkteal,
      fill=darkteal,
  }}

  \begin{subfigure}[b]{0.24\linewidth}
    \begin{tikzpicture}
      \begin{axis}
        [
          param1,
          xtick={1, 2, 3, 4},
          xticklabels={4\%, 10\%, 20\%, 40\%},
          xmin=0.5,
          xmax=4.5,
          xlabel=ratio of frequently accessed
          columns
        ]

        \addplot[parambar]
        table[x=x,y=y]{
          x	y
          1	68.9
          2	74.8
          3	41.7
          4	28.0
        };

      \end{axis}
    \end{tikzpicture}
    \caption{Workload Diversity}
  \end{subfigure}
  ~
  \begin{subfigure}[b]{0.24\linewidth}
    \begin{tikzpicture}
      \begin{axis}
        [
          param1,
          xtick={1, 2, 3},
          xticklabels={Uniform, Gaussian, Skewed},
          xmin=0.5,
          xmax=3.5,
          x tick label style={rotate=30, anchor=east},
        ]

        \addplot[parambar]
        table[x=x,y=y]{
          x	y
          1	69.8
          2	61.3
          3	64.4
        };

      \end{axis}
    \end{tikzpicture}
    \caption{Data distribution}
  \end{subfigure}
  ~
  \begin{subfigure}[b]{0.24\linewidth}
    \begin{tikzpicture}
      \begin{axis}
        [
          width=45mm,
          height=30mm,
          xtick={10, 100, 200, 300, 400},
          draw=darkteal,
          fill=none,
          xmin=10,
          xmax=400,
          ymin=0,
          ymax=0.805,
          ytick={0, 0.2, ..., 0.8},
          yticklabels={0, 20, 40, 60, 80},
          xlabel=\# of past queries,
          ylabel near ticks,
          ylabel shift=-2pt,
          xlabel near ticks,
          ylabel style={align=center},
          ylabel=Error\\ Reduction (\%),
        ]

        \addplot[parambar,
          fill=none,
          ultra thick,
        ]
        table[x=x,y=y]{
          x	y
           10	0.134
           20	0.433
           30	0.457
           40	0.485
           50	0.508
           60	0.539
           70	0.561
           80	0.595
           90	0.610
           100	0.620
           110	0.635
           120	0.640
           130	0.647
           140	0.650
           150	0.658
           160	0.669
           170	0.672
           180	0.679
           190	0.629
           200	0.635
           210	0.603
           220	0.614
           230	0.616
           240	0.625
           250	0.632
           260	0.634
           270	0.629
           280	0.638
           290	0.638
           300	0.638
           310	0.638
           320	0.644
           330	0.652
           340	0.617
           350	0.618
           360	0.615
           370	0.616
           380	0.620
           390	0.612
           400	0.599
        };

      \end{axis}
    \end{tikzpicture}
    \caption{Learning Behavior}
  \end{subfigure}
  ~
  \begin{subfigure}[b]{0.24\linewidth}
    \begin{tikzpicture}
      \begin{axis}
        [
          width=45mm,
          height=30mm,
          xtick={10, 100, 200, 300, 400},
          draw=darkteal,
          fill=none,
          xmin=10,
          xmax=400,
          ymin=0,
          ymax=10,
          ytick={0, 2, ..., 10},
          xlabel=\# of past queries,
          ylabel near ticks,
          ylabel shift=-2pt,
          xlabel near ticks,
          ylabel style={align=center},
          ylabel=Overhead (ms),
        ]

        \addplot[parambar,
          fill=none,
          ultra thick,
          ] table[x=x,y=y]{
          x	y
          10	1
          20	1
          30	1
          40	0
          50	0
          60	0
          70	1
          80	0
          90	0
          100	1
          110	1
          120	1
          130	1
          140	0
          150	0
          160	0
          170	0
          180	1
          190	1
          200	1
          210	1
          220	2
          230	1
          240	1
          250	0
          260	0
          270	1
          280	0
          290	1
          300	0
          310	1
          320	0
          330	1
          340	1
          350	1
          360	0
          370	1
          380	1
          390	1
          400	1
        };

      \end{axis}
    \end{tikzpicture}
    \caption{Overhead}
  \end{subfigure}

  \caption{The effectiveness of \dbl  in reducing \nol's error for different (a) levels of diversity in the queried columns, 
     (b) data   distributions, and (c) number of past queries observed. Figure
    (d) shows \dbl's overhead for different number of past queries.}
  \label{fig:exp:param}
  \vspace{\neggap}
\end{figure*}




\subsection{Impact of Different Data Distributions and Workload Characteristics}
\label{sec:exp:parameter}

In this section, we generated various synthetic datasets and queries to fully
understand how \dbl's  effectiveness changes for different data distributions,
query patterns, and number of past queries.

First, we studied the impact of having queries with a  more diverse  set of columns in their selection predicates. 
We produced
a table of 50 columns and 5M rows,
\rev{C6}{where 10\% of the columns were categorical. The domains of the numeric columns were the real values
between 0 and 10, and the domains of the categorical columns were the integers
between 0 and 100.}

Also, we
generated four different query workloads
with varying proportions of frequently accessed columns. 
The columns used
for the selection predicates were chosen according to a power-law
distribution. Specifically,
     a fixed number of columns (called \emph{frequently accessed columns}) had
      the same probability of being accessed, but the access probability of the remaining columns decayed according to the power-law distribution.
For instance, if the proportion of frequently accessed columns was 20\%,
the first 20\% of the columns (\ie, 10 columns) appeared with equal probability in each query, 
but the probability of appearance reduced by half for every remaining column. 
  \Cref{fig:exp:param}(a) shows that as the proportion of frequently
accessed columns increased, \dbl's relative error reduction over \nol gradually
decreased (the number of past queries were fixed to 100). This is expected 
as \dbl constructs its
model based on the
columns appearing in the past.
In other words,  to cope with  the increased diversity, more past queries are needed to understand the complex
underlying distribution that generated the data. Note that, according to
the analytic queries in the \vertica dataset, most of the queries included less
than 5 distinct selection predicates. 
However, by processing  more queries,
\dbl continued to learn more about the underlying distribution, and produced
larger error reductions even when the workload was extremely diverse (\cref{fig:exp:param}(c)).

Second, to study \dbl's potential sensitivity,
 we generated three
tables using three different probability distributions: uniform,
 Gaussian, and a log-normal (skewed) distribution. 
 \Cref{fig:exp:param}(b) shows \dbl's error
 reductions when queries were run
 against  each table. 
 Because of the power and generality of the maximum entropy principle taken by
 \dbl, it delivered a consistent performance irrespective of the underlying
distribution.


Third, we varied the number of past queries  observed by \dbl before running our
test queries. For this study, we used a highly diverse query set (its
proportion of frequently accessed columns was 20\%).
\Cref{fig:exp:param}(c) demonstrates that the error
reduction continued increasing as more queries were processed, but its increment slowed down.
This is because, after observing enough information, \dbl already had a good knowledge of the
underlying distribution, and processing more queries barely improved its knowledge. This result
indicates that \dbl is able to deliver reasonable performance without having to observe too many
queries.

This is because, after observing enough information, \dbl already has a good knowledge of the
underlying distribution, and processing more queries barely improves its knowledge. This result
indicates that \dbl is able to deliver reasonable performance without having to observe too many
queries.


Lastly, we studied the negative impact of increasing the number of past queries on  \dbl's overhead.
Since \dbl's inference consists of a small matrix multiplication,
we did not observe a noticeable increase in its runtime overhead even
when the number of queries in the
query synopsis increased (Figure~\ref{fig:exp:param}(d)).

\rev{C6}{Recall that the domain size of the attributes does not affect \dbl's
computational cost since only the lower and upper bounds of range
constraints are needed for covariance computations (\cref{sec:data_stat}).}

\section{Related Work}
\label{sec:related}



\ph{Approximate Query Processing}
There has been substantial work on  sampling-based  approximate query
processing~\cite{surajit-optimized-stratified,dynamicp-sample-selection,join_synopses,aqua2,interactive-cidr,sciborq,mozafari_eurosys2013,ganti2000icicles,hose2006distributed,kandula2016quickr,considine2004approximate,meliou2009approximating,potti2015daq,wang2014sample,fan2015querying}. Some of these systems differ in their sample generation strategies. 
Some of these systems differ in their sample generation strategies (see~\cite{approx_chapter} and the references within). 
 For instance, STRAT~\cite{surajit-optimized-stratified} and
AQUA~\cite{aqua2} create a single stratified sample, while BlinkDB creates samples based on \emph{column sets}.
Online Aggregation (OLA)~\cite{online-agg,online-agg-mr2,online-agg-mr1,cosmos} continuously
refines its answers during query execution. 
Others  have focused on obtaining faster or more reliable error estimates~\cite{xu2008confidence,mozafari_sigmod2014_diagnosis}.
These are orthogonal to our work, as reliable  error estimates from an underlying AQP engine will also benefit DBL.
%
There is also AQP techniques developed for specific domain, e.g., sequential
data~\cite{arasu2004approximate,perera2016efficient},
probabilistic
data~\cite{gatterbauer2015approximate,olteanu2010approximate},
and RDF data~\cite{huang2012approximating,souihli2013optimizing}, and
searching in high-dimensional space~\cite{mozafari_pvldb2015_ksh}.
However, our focus in this paper is on general (SQL-based) AQP engines.

\ph{Adaptive Indexing, View Selection}
Adaptive Indexing and database
cracking~\cite{db-cracking,holistic-indexing,idreos2009self} has been proposed
for a column-store database as a means of incrementally updating
indices as part of query processing; then, it can speed up future queries that
access previously indexed ranges.
\rev{C15}{While the adaptive indexing is an effective mechanism for exact analytic
query processing in column-store databases, answering queries that require
accessing multiple columns (\eg, selection predicates on multiple columns)
is still a challenging task: column-store
databases have to join relevant columns to reconstruct tuples.
Although Idreos \etal~\cite{idreos2009self}
pre-join some subsets of columns, the number of column combinations still grows exponentially
as the total number of columns in a table increases. 
 \dbl can easily handle 
queries with multiple columns due to its analytic inference.}
Materialized views are another means of speeding up future
aggregate queries~\cite{joshi2008materialized,halevy2001answering,el2009statadvisor,mozafari_sigmod2015}.
\dbl also speed up aggregate queries, but \dbl 
  does not require strict query containments as in materialized views.

\ph{Pre-computation} COSMOS~\cite{cosmos}  
  stores the results of past queries as multi-dimensional cubes,
  which are then reused
   if they are contained in the new query's input range, 
  	while boundary tuples  are read from the database. 
This approach is not probabilistic and is limited to low-dimensional data 	
 due to the exponential explosion in the number of possible cubes.
Also, similar to view selection, COSMOS relies on strict query containment.

\ph{Model-based and Statistical Databases}
Statistical approaches have been used in databases for various goals. 
MauveDB~\cite{deshpande2006mauvedb} constructs views that express a statistical
model, hiding the possible irregularities of the underlying data.
MauveDB's goal is to
support statistical modeling, such as regression or interpolation.
, rather than
speeding up future query processing.
BayesDB~\cite{mansinghka2015bayesdb} provides a SQL-like language that enables
non-statisticians to declaratively use various statistical models.
\rev{C11}{
Bayesian networks have been used for succinctly capturing
correlations among attributes~\cite{getoor2001selectivity}.
Exploiting these correlations can be an interesting future direction for DBL.}

\ph{Maximum Entropy Principle}
\rev{C11}{
In the database community, the principle of maximum entropy (ME) has been previously
used for determining the most surprising piece of
information in a data exploration context~\cite{sarawagi2000user}, and 
for constructing histograms based on cardinality
assertions~\cite{kaushik2009general}.
\dbl uses ME differently than these previous approaches;
they assign a unique variable to each non-overlapping area to
represent the number of tuples belonging to that area. This approach poses
two challenges when applied to an AQP context. First, it requires a slow
iterative numeric solver for its inference.
Thus, using this approach for DBL may eliminate any potential speedup.
Second, introducing a unique variable for every non-overlapping area can be
impractical as it requires  $O(2^n)$ variables for $n$ past queries.
Finally, the previous approach cannot express inter-tuple covariances in
the underlying data.
In contrast,
\dbl's approach handles arbitrarily overlapping ranges in multidimensional space
with $O(n)$ variables (and $O(n^2)$ space), and its inference can be performed analytically.
}



\section{Conclusion and Future Work}
\label{sec:con}

In this paper, we presented database learning (DBL), a novel approach to exploit past queries'
(approximate) answers in speeding up new queries using a principled statistical methodology.
We presented a prototype of this vision, called \dbl, on top of Spark SQL.
Through extensive experiments on real-world and benchmark query logs, 
we demonstrated that \dbl supported 73.7\% of real-world analytical queries,
speeding them up  by up to 23$\times$ compared to an online aggregation AQP engine.


Exciting lines of future work include: (i) the study of other inferential
techniques for realizing DBL, 
(ii) the development of \emph{active database learning}~\cite{park2017active}, whereby the engine
itself proactively executes  certain approximate queries 
that can best improve its internal model, and (iii) the extension
of \dbl to support visual
analytics~\cite{el2016vistrees,park2016visualization,crotty2015vizdom,joglekar2014interactive,kim2015rapid,vartakRMPP15,battle2015dynamic,wu2014case,idreos2015overview}.

\section{Acknowledgement}

This work was in part funded by NSF awards 1544844, 1629397, and 1553169.
The authors are grateful to Aditya Modi, Morgan Lovay, and Shannon Riggins for their 
	feedback on this manuscript.






%

\small
\bibliographystyle{abbrv}
\bibliography{ldbbib}  
\normalsize

%
%

\appendix

Appendix is organized as follows.
\Cref{sec:learn,sec:safeguard} provide \dbl's learning and model validation
processes, respectively.
\Cref{sec:more:exp} includes additional experiments for demonstrating the effectiveness and
    accuracy of \dbl.
\Cref{sec:append} present the generalization of \dbl's approach to the databases
into which new tuples may be inserted.
\Cref{sec:covariance:study} studies the prevalence of non-zero inter-tuple
correlations in real-world datasets.
\Cref{sec:tech_details} describes technical details that we did not discuss in
the main body of the paper.



%

\section{Parameter Learning}
\label{sec:learn}

This section describes \dbl's correlation parameter learning. In
\cref{sec:learn:optimal}, we presents mathematical description of the process,
and in \cref{sec:exp:correlation}, we study its effectiveness with experiments.

\subsection{Optimal Correlation Parameters}
\label{sec:learn:optimal}


In this section, we describe how to find the most likely values
of the correlation parameters defined in \cref{sec:data_stat}.
In this process, we exploit the joint pdf defined in \cref{eq:prior_pdf}, as it
	allows us to compute the
		likelihood of a certain combination of query answers given relevant statistics.
 Let
$\estansvec_{\text{past}}$ denote a vector of raw answers to past snippets. Then,
by Bayes' theorem:
\[
  \pr(\Sigma_n \mid \estansvec_{\text{past}})
  \propto \pr(\Sigma_n) \cdot \pr(\estansvec_{\text{past}} \mid \Sigma_n)
\]
where
\rev{C2.1e}{$\Sigma_n$ is an $n \times n$ submatrix of $\Sigma$ consisting of
$\Sigma$'s first $n$ rows and columns, \ie, (co)variances between pairs of past
query answers,}
and $\propto$ indicates that the two values are proportional,
Therefore, without any preference over parameter values, 
determining the most likely correlation parameters
(which determine $\Sigma_n$) given past queries
 amounts to
finding the values for $l_{g,1}, \ldots, l_{g,l}, \allowbreak \sigma_g^2$ that maximize the
below log-likelihood function:
\begin{align}
  &\log \, \pr(\estansvec_{\text{past}} \mid \Sigma_n)
  = \log \, f(\estansvec_{\text{past}}) \nonumber \\
  &= - \frac{1}{2} \estansvec_{\text{past}}^\intercal \Sigma_n^{-1} \estansvec_{\text{past}}
  - \frac{1}{2} \log |\Sigma_n| - \frac{n}{2} \log 2 \pi
  \label{eq:log_likelihood}
\end{align}
where $f(\estansvec_{\text{past}})$ is the joint pdf from 
\cref{eq:prior_pdf}.

\rev{C2.1d}{
\dbl finds the optimal values for $l_{g,1}, \ldots, l_{g,l}$ by solving the
above optimization problem with a numerical solver, while it estimates the value
for $\sigma_g^2$ analytically from past query answers (see \cref{sec:anal_param}).
Concretely,
the current implementation of \dbl uses the gradient-descent-based
(quasi-newton) nonlinear
programming solver provided by Matlab's \texttt{fminuncon()} function,
without providing explicit gradients.}
Although our current approach is typically slower than
using closed-form solutions or than using the solver with an explicit gradient (and
a Hessian), 
they do not pose a challenge in \dbl's setting, since
 all these parameters are computed \emph{offline}, \ie, prior to the arrival of new queries.
\rev{C2.1d}{We plan to improve the efficiency of this offline training by using
explicit gradient expressions.}

\begin{figure}[t]

  \pgfplotsset{corranal/.style={
      width=70mm,
      height=35mm,
      xmin=0,
      xmax=1,
      ymin=0,
      ymax=1.1,
      ytick={0, 0.2, ..., 1.0},
      ylabel near ticks,
      xlabel near ticks,
      ylabel style={align=center},
      xlabel=True Correlation Parameter,
      ylabel=Estimated\\ Correlation\\ Parameter,
      legend style={
        at={(0.37,1)},anchor=south west,column sep=3pt,
        draw=none,font=\scriptsize,fill=none,font=\scriptsize,line width=.5pt,
        /tikz/every even column/.append style={column sep=5pt}
      },
      legend columns=3,
      every axis/.append style={font=\scriptsize},
      tick style={draw=none},
      legend cell align=left,
      xmajorgrids,
      ymajorgrids,
  }}

  \pgfplotsset{scatterdots/.style={
      fill=gray,
      draw=none,
      only marks,
      mark size=1.5,
      opacity=0.8,
  }}

  \centering
  \begin{tikzpicture}
    \begin{axis}[corranal]

      \addplot[scatterdots]
      table[x=x,y=y1] {data/cov_analysis.csv};

      \addplot[scatterdots,fill=darkteal]
      table[x=x,y=y2] {data/cov_analysis.csv};

      \addplot[scatterdots,fill=darkorange]
      table[x=x,y=y3] {data/cov_analysis.csv};

      \addplot[draw=black,mark=none] coordinates {(0,0) (1,1)};


      \addlegendentry{20};
      \addlegendentry{50};
      \addlegendentry{100};
    \end{axis}

    \node[font=\scriptsize] at (0.7,2.2) {\# past queries:};

  \end{tikzpicture}

  \caption{Correlation Param Learning}
  \label{fig:param_learning}
\end{figure}
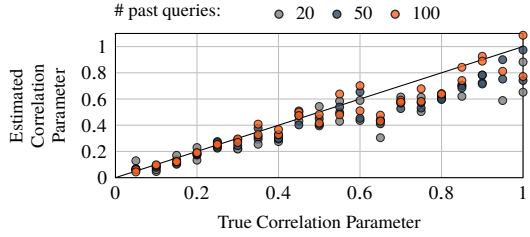

\rev{C2.1d}{
Since \cref{eq:log_likelihood} is not a convex function, the solver of our
choice only returns a locally optimal point. A conventional strategy to handle
this issue is to obtain multiple locally optimal points by solving
the same problem with multiple random starting points, and
to take the one with the highest log-likelihood value as a final answer.
Still, this approach does not guarantee the correctness of the model.
In contrast,
\dbl's strategy is to find a locally-optimal point that can capture potentially large inter-tuple
covariances, and to validate the correctness of the resulting model against a
model-free answer (\cref{sec:safeguard}).
We demonstrate empirically in the following section that this strategy is
effective for finding parameter values that are close to true values.
\dbl's model validation process in \cref{sec:safeguard} provides robustness
against the models that may differ from the true distribution.
\dbl uses $l_{g,k} = (\text{max}(A_k) - \text{min}(A_k))$ for the starting point
of the optimization problem.}

Lastly, our use of approximate
answers as the constraints for the ME principle is properly accounted for
 by including additive error terms in their (co)variances
(\cref{eq:query_cov2}).


\subsection{Accuracy of Parameter Learning}
\label{sec:exp:correlation}

\afterrev{
In this section, we demonstrate our empirical study on
the effectiveness of \dbl's correlation parameter estimation process.
For this, we used the synthetic datasets
generated from pre-determined correlation parameters to see how close \dbl could estimate the values
of those correlation parameters. We let \dbl estimate the correlation parameter values using three
different numbers of past snippets (20, 50, and 100) for various datasets with different
correlation parameter values.}

\afterrev{
\Cref{fig:param_learning} shows the results. In general, the correlation parameter values discovered
by \dbl's estimation process were consistent with the true correlation parameter values. Also, the
estimated values tended to be closer to the true values when a larger number of past snippets were
used for the estimation process. This result indicates that \dbl can effectively learn statistical
characteristics of the underlying distribution just based on the answers to the
past queries.}


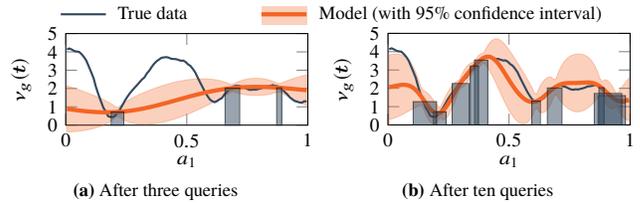
\begin{figure}[t]
  \centering

  \pgfplotsset{axisIntroFigure/.style={
          width=48mm,
          height=28mm,
          xmin=0,
          xmax=1,
          ymin=0,
          ymax=5,
          legend style={at={(-0.06,1)},anchor=south west,
            column sep=2pt,
            /tikz/every even column/.append style={column sep=25.2pt},
          draw=none,font=\scriptsize,fill=none,line width=.5pt},
          legend cell align=left,
          legend columns=2,
          clip=false,
          ytick={0, 1, 2, 3, 4, 5},
          xtick={0, 0.5, 1.0},
          ylabel=$\nu_g(\mtuple)$,
          xlabel=$\tcol_1$,
          ylabel near ticks,
          xlabel near ticks,
          xlabel shift=-1mm,
  }}

  \pgfplotsset{curve/.style={%
          mark=none,
          opacity=1.0
  }}

  \tikzset{barstyle/.style={%
          opacity=0.5,
          fill=darkteal,
  }}

  \tikzset{legendnode/.style={%
          opacity=0.5,
          fill=tomato,
          minimum width=5mm,
          minimum height=2.5mm,
          anchor=south west,
  }}

  \begin{subfigure}[b]{0.48\linewidth}
    \centering
    \begin{tikzpicture}
      \begin{axis}[axisIntroFigure]

        \addplot[curve,tomato,fill=tomato,opacity=0.3,forget plot] table [x=x, y=a, col sep=tab] {learn_fig_data2.txt} \closedcycle;    
        \addplot[curve,thick,darkteal] table [x=x, y=y, col sep=tab] {learn_fig_data1.txt};            
        \addplot[curve,tomato,ultra thick] table [x=x, y=a, col sep=tab] {learn_fig_data1.txt};    

        \addlegendentry{True data}

        \addlegendentry{Model (with 95\% confidence interval)}

        \draw [barstyle] (axis cs:0.658,0) rectangle (axis cs:0.719,2.022);
        \draw [barstyle] (axis cs:0.187,0) rectangle (axis cs:0.24,0.719);
        \draw [barstyle] (axis cs:0.871,0) rectangle (axis cs:0.892,2.04);


        \node [legendnode,tomato,opacity=0.3,minimum width=6mm,minimum height=2.5mm]
        at (axis cs:0.81, 5.45) {};

      \end{axis}
    \end{tikzpicture}
    \vspace{-5mm}
    \caption{After three queries}
  \end{subfigure}
  ~
  \begin{subfigure}[b]{0.48\linewidth}
    \centering
    \begin{tikzpicture}
      \begin{axis}[axisIntroFigure,clip=true]

        \addplot[curve,tomato,fill=tomato,opacity=0.3,forget plot] table [x=x, y=b, col sep=tab] {learn_fig_data2.txt} \closedcycle;    
        \addplot[curve,thick,darkteal] table [x=x, y=y, col sep=tab,forget plot] {learn_fig_data1.txt};            
        \addplot[curve,tomato,ultra thick] table [x=x, y=b, col sep=tab] {learn_fig_data1.txt};    


        \draw [barstyle] (axis cs:0.658,0) rectangle (axis cs:0.719,2.022);
        \draw [barstyle] (axis cs:0.187,0) rectangle (axis cs:0.24,0.719);
        \draw [barstyle] (axis cs:0.871,0) rectangle (axis cs:0.892,2.04);
        \draw [barstyle] (axis cs:0.867,0) rectangle (axis cs:0.981,1.606);
        \draw [barstyle] (axis cs:0.852,0) rectangle (axis cs:0.966,1.725);
        \draw [barstyle] (axis cs:0.357,0) rectangle (axis cs:0.412,3.544);
        \draw [barstyle] (axis cs:0.265,0) rectangle (axis cs:0.335,2.269);
        \draw [barstyle] (axis cs:0.104,0) rectangle (axis cs:0.201,1.275);
        \draw [barstyle] (axis cs:0.594,0) rectangle (axis cs:0.630,1.304);
        \draw [barstyle] (axis cs:0.34 ,0) rectangle (axis cs:0.374,3.224);

      \end{axis}
    \end{tikzpicture}
    \vspace{-5mm}
    \caption{After ten queries}
  \end{subfigure}

  \caption{
    An example of (a) overly optimistic confidence intervals due to incorrect estimation of the
    underlying distributon, and (b) its resolution with more queries processed.  \dbl relies on a
    model validation to avoid the situation as in (a).
  }
    
  \label{fig:learn:safeguard}
  \vspace{\neggap}
\end{figure}

\section{Model Validation}
\label{sec:safeguard}



\dbl aims to provide correct error bounds even when its model
differs significantly from the true data. In \cref{sec:safeguard:process}, we
describes its process, and in \cref{sec:exp:validation}, we empirically demonstrate
its effectiveness.

\subsection{Model Validation Procedure}
\label{sec:safeguard:process}

\dbl'd model validation
rejects its model---the most likely explanation of the underlying distribution
given the answers to past snippets---if there is evidence that its
model-based error is likely to be incorrect.
\dbl's model validation process addresses two situations: (i) negative estimates for \texttt{FREQ(*)},
and (ii) an unlikely large discrepancy between a model-based answer and a raw
answer.

\ph{Negative estimates for \texttt{FREQ(*)}}
\rev{C12}{
To obtain the prior distribution of the random variables,
\dbl uses the most-likely distribution (based on the maximum entropy
principle (\cref{lemma:pdf})), given the means,
variances, and covariances of query answers. Although this makes
the inference analytically computable, the lack of explicit non-negative
constraints on the query answers may produce negative estimates on
\texttt{FREQ(*)}. \dbl handles this situation with a simple check; that is,
\dbl rejects its model-based answer if $\mans < 0$ for \texttt{FREQ(*)}, and
uses the raw answer instead. Even if $\mans \ge 0$, the lower bound of the
confidence interval is set to zero if the value is less than zero.
}



\ph{Unlikely model-based answer}
\dbl's model learned from empirical observations may be different from the true
distribution.
\Cref{fig:learn:safeguard}(a) illustrates such an example.
Here, after the first three queries,
	 the model is consistent with past query answers (shown as gray boxes);
however, it incorrectly estimates the distribution of the unobserved data,
	leading to overly optimistic confidence
intervals. 
\Cref{fig:learn:safeguard}(b) shows that the model becomes more consistent with the data as
	more queries are processed.

  \dbl rejects (and does not use) its own model in situations such as \cref{fig:learn:safeguard}(a)
 by validating its model-based answers against the \emph{model-free} answers obtained from the AQP engine. Specifically,
 	we define a \emph{likely region} as the range in which 
	the AQP
		engine's answer would fall with high probability (99\% by default) if \dbl's
		model were to be correct.
	If the AQP's raw answer $\estans_{n+1}$ falls outside this likely
		region, \dbl considers its model unlikely to be correct.
	In such cases, \dbl drops its model-based answer/error, and simply returns the raw
		answer to the user unchanged. This process is akin to hypothesis testing
		in statistics literatures~\cite{freedman2007statistics}.


Although no improvements are
	made in such cases, we take this conservative approach to
	 ensure the correctness of our error guarantees.
  (See \cref{sec:benefit,sec:exp:validation}
for formal guarantees and empirical results, respectively).

Formally, let $t\geq 0$ be the value for which 
	 the AQP
engine's answer would fall within 
$(\mans - t,\, \mans + t)$
with probability $\delta_v$ (0.99 by default) if $\mans$ were the
exact answer.
We call the $(\mans - t,\, \mans + t)$ range the likely region.
To compute $t$,
we must find the value closest to $\mans$ that satisfies the
\rev{C2.3}{following expression}:
\begin{align}
  \pr\left( | X - \mans | < t \right) \ge \delta_v
  \label{eq:model_val}
\end{align}
where $X$ is a random variable representing the AQP engine's possible answer to 
  the new snippet if the exact answer to the new snippet was
$\mans$. The AQP engine's answer can be treated as a random variable
since it may differ depending on the random samples used.
The probability in \cref{eq:model_val} can be easily
computed using either the central limit theorem or the Chebyshev's
inequality~\cite{lovric2011international}. 
Once the value of $t$ is computed, \dbl rejects its model if $\rawans_{n+1}$ falls outside the likely region $(\mans - t,\, \mans + t)$.

In summary, the pair of \dbl's improved answer and improved error, $(\impmean, \impstd)$, is
set to $(\mans, \mstd)$ if $\rawans_{n+1}$ is within the range $(\mans - t, \mans + t)$,
and is set to $(\estans_{n+1}, \beta_{n+1})$ otherwise. 
 In either case, the
error bound at confidence $\delta$ remains the same
as $\alpha_\delta \cdot \impstd$,
where $\alpha_\delta$ is the confidence interval multiplier for probability
$\delta$.

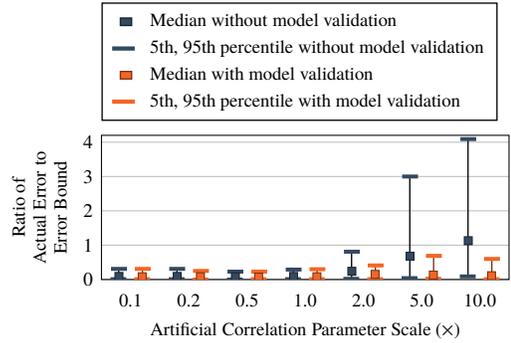
\begin{figure}[t]
  \pgfplotsset{confsize/.style={
      width=70mm,
      height=35mm,
      xmin=0.5,
      xmax=7.5,
      ymin=0,
      ymax=4.2,
      ybar=4pt,
      bar width=5pt,
      xtick={1, 2, ..., 7},
      xticklabels={0.1, 0.2, 0.5, 1.0, 2.0, 5.0, 10.0},
      ytick={0, 1, 2, 3, 4},
      ylabel near ticks,
      ylabel style={align=center},
      xlabel=Artificial Correlation Parameter Scale ($\times$),
      ylabel=Ratio of\\ Actual Error to \\ Error Bound,
      legend style={
        at={(0.0,1.1)},anchor=south west,column sep=3pt,
        draw=black,font=\scriptsize,fill=white,line width=.5pt,
      },
      legend columns=1,
      every axis/.append style={font=\scriptsize},
      tick style={draw=none},
      ymajorgrids,
      legend cell align=left,
      x tick label style={yshift=2pt},
  }}

  \pgfplotsset{graybar/.style={
      fill=none,
      draw=none,
      error bars/.cd,
      y dir=both,
      y explicit,
      error bar style={line width=0.5pt,black},
      error mark options={
          rotate=90,
          darkteal,
          mark size=3pt,
          line width=1.5pt
      },
  }}

  \pgfplotsset{redbar/.style={
      fill=none,
      draw=none,
      error bars/.cd,
      y dir=both,
      y explicit,
      error bar style={line width=0.5pt,black!50!darkorange},
      error mark options={
          rotate=90,
          darkorange,
          mark size=3pt,
          line width=1.5pt
      },
  }}

  \tikzset{mediandot/.style={
    scale=0.5,fill=darkteal,draw=black!50!darkteal,rectangle,
  }}

  \pgfplotsset{
    box legend image/.style={
      legend image code/.code={%
        \node[minimum width=1.5mm,minimum height=1mm,
              fill=#1,draw=black!50!#1,inner sep=0]
        at (0.2cm,0cm) {};
      }
    },
  }

  \pgfplotsset{
    errorbar legend image/.style={
      legend image code/.code={%
        \draw[ultra thick,#1] (0.1cm,0cm)--(0.4cm,0cm);
      }
    },
  }

  \centering
  \begin{tikzpicture}
    \begin{axis}[confsize]
      \addlegendimage{box legend image=darkteal};
      \addlegendentry{Median without model validation};

      \addlegendimage{errorbar legend image=darkteal};
      \addlegendentry{5th, 95th percentile without model validation};

      \addlegendimage{box legend image=darkorange};
      \addlegendentry{Median with model validation};

      \addlegendimage{errorbar legend image=darkorange};
      \addlegendentry{5th, 95th percentile with model validation};

      \addplot[graybar]
      table [y error plus=z2, y error minus=z1] {
        x	y	z1	z2
        1	0.09	0.08	0.22
        2	0.09	0.08	0.22
        3	0.08	0.08	0.15
        4	0.09	0.08	0.20
        5	0.24	0.22	0.57
        6	0.68	0.64	2.32
        7	1.13	1.04	2.96
      };

      \addplot[redbar]
      table [y error plus=z2, y error minus=z1] {
        x	y	z1	z2
        1	0.08	0.07	0.23
        2	0.08	0.08	0.17
        3	0.07	0.06	0.16
        4	0.08	0.07	0.22
        5	0.15	0.14	0.26
        6	0.13	0.11	0.56
        7	0.11	0.10	0.49
      };

      \node[mediandot] at (axis cs:0.8, 0.09) {};
      \node[mediandot] at (axis cs:1.8, 0.09) {};
      \node[mediandot] at (axis cs:2.8, 0.08) {};
      \node[mediandot] at (axis cs:3.8, 0.09) {};
      \node[mediandot] at (axis cs:4.8, 0.24) {};
      \node[mediandot] at (axis cs:5.8, 0.68) {};
      \node[mediandot] at (axis cs:6.8, 1.13) {};

      \node[mediandot,fill=darkorange,draw=black!50!darkorange] at (axis cs:1.2, 0.08) {};
      \node[mediandot,fill=darkorange,draw=black!50!darkorange] at (axis cs:2.2, 0.08) {};
      \node[mediandot,fill=darkorange,draw=black!50!darkorange] at (axis cs:3.2, 0.07) {};
      \node[mediandot,fill=darkorange,draw=black!50!darkorange] at (axis cs:4.2, 0.08) {};
      \node[mediandot,fill=darkorange,draw=black!50!darkorange] at (axis cs:5.2, 0.15) {};
      \node[mediandot,fill=darkorange,draw=black!50!darkorange] at (axis cs:6.2, 0.13) {};
      \node[mediandot,fill=darkorange,draw=black!50!darkorange] at (axis cs:7.2, 0.11) {};


    \end{axis}

  \end{tikzpicture}

  \caption{Effect of model validation. For \dbl's error bounds to be correct, the 95th percentile
  should be below 1.0. One can find that, with \dbl's model validation, the improved answers and the
  improved errors were probabilistically correct even when largely incorrect correlation parameters
  were used.}
  \label{fig:model_validation}
\end{figure}

\subsection{Empirical Study on Model Validation}
\label{sec:exp:validation}

This section studies the effect of \dbl's model validation described in \cref{sec:safeguard:process}. For
this study, we first generated synthetic datasets with several predetermined correlation parameters
values.
Note that one can generate such synthetic datasets by first determining a joint probabilistic
distribution function with predetermined correlation parameter values and sampling attribute values from the
joint probability distribution function. In usual usage scenario, \dbl estimates those correlation
parameters from past snippets; however, in this section, we manually set the values for the
correlation parameters in \dbl's model, to test the behavior of \dbl running with possibly incorrect
correlation parameter values.

\Cref{fig:model_validation} reports the experiment results from when \dbl was tested without a model
validation step and with a model validation step, respectively. In the figure,
the values on the X-axis are artificial correlation parameter scales, \ie, the product of the true correlation
parameters and each of those scales are set in \dbl's model.
For instance, if a true correlation parameter was 5.0, and the ``artificial correlation
parameter scale'' was 0.2, \dbl's model was set to 1.0 for the correlation parameter. Thus, the values of
the correlation parameters in \dbl's model were set to the true correlation parameters, when
the ``artificial correlation
parameter scale'' was 1.0. Since the Y-axis reports the ratio of the actual error to \dbl's error
bound, \dbl's error bound was correct when the value on the Y-axis was below 1.0.

In the figure, one can observe that, \dbl, used without the model validation, produced incorrect error
bounds when the correlation parameters used for the model deviated largely from the true correlation
parameter values.
However, \dbl's model validation could successfully identify incorrect model-based answers and
provide correct error bounds by replacing those incorrect model-based answers with the raw answers
computed by the AQP system.

\section{Additional Experiments}
\label{sec:more:exp}

This section contains additional experiments we have not included in the main part of our paper.
  First, we study the impact of two factors that affect the performance benefits of \dbl
    (\cref{sec:exp:contribution}).
  Second, we show that \dbl can achieve error reductions over time-bounded AQP
  engines (\cref{sec:time-bound}).


\subsection{\dbl vs.~Simple Answer Caching}
\label{sec:exp:contribution}

\begin{figure}[t]
  \begin{subfigure}[b]{1.0\linewidth}
    \pgfplotsset{pastaccgraph/.style={
        width=65mm,
        height=32mm,
        xtick={1, 2, ..., 5},
        xticklabels={0.01, 0.05, 0.1, 0.5, 1.0},
        ytick={0,20,40,60,80,100},
        xmin=0.5,
        xmax=5.5,
        ymin=0,
        ymax=100,
        xlabel near ticks,
        ybar,
        bar width=5pt,
        xlabel near ticks,
        xlabel style={align=center},
        xlabel=Sample Size (\%),
        ylabel=Actual Error\\ Reduction (\%),
        ylabel style={align=center},
        legend style={
          at={(0,1.1)},anchor=south west,column sep=2pt,
          draw=none,font=\scriptsize,fill=none,line width=.5pt,
          /tikz/every even column/.append style={column sep=10pt}
        },
        legend columns=2,
        every axis/.append style={font=\scriptsize},
        area legend,
        ymajorgrids,
        xtick style={draw=none},
        x tick label style={yshift=5pt},
        legend cell align=left,
    }}

    \pgfplotsset{tealbar/.style={
        draw=black!50!darkteal,
        fill=darkteal,
    }}

    \pgfplotsset{redbar/.style={
        draw=black!50!darkorange,
        fill=darkorange,
        postaction={
          pattern=north west lines,
        }
    }}

    \centering
    \begin{tikzpicture}
      \begin{axis}[pastaccgraph]

        \addplot[tealbar]   
        table[x=x,y=y]{
          x	y
        1	23.044
        2	34.141
        3	35.904
        4	40.104
        5	40.052
        };

        \addplot[redbar]   
        table[x=x,y=y]{
          x	y
        1	53.127
        2	74.063
        3	77.650
        4	89.802
        5	85.541
        };

        \addlegendentry{\basel};
        \addlegendentry{\dbl};

      \end{axis}
    \end{tikzpicture}

    \caption{Sample Sizes for Past Queries}
  \end{subfigure}

  \vspace{2mm}

  \begin{subfigure}[b]{1.0\linewidth}
    \pgfplotsset{contgraph/.style={
        width=65mm,
        height=32mm,
        ybar,
        xtick={1, 3, 5, 7, 9, 11},
        xticklabels={0, 20, 40, 60, 80, 100},
        ytick={0, 20, ..., 100},
        xmin=0,
        xmax=12,
        ymin=0,
        ymax=100,
        bar width=5pt,
        xlabel near ticks,
        xlabel near ticks,
        xlabel style={align=center},
        xlabel=Novel Query Ratio (\%),
        ylabel=Actual Error\\ Reduction (\%),
        ylabel style={align=center},
        legend style={
          at={(0,1)},anchor=south west,column sep=2pt,
          draw=none,font=\tiny,fill=none,line width=.5pt,
          /tikz/every even column/.append style={column sep=10pt}
        },
        legend columns=1,
        every axis/.append style={font=\scriptsize},
        area legend,
        xtick style={draw=none},
        x tick label style={yshift=5pt},
        ymajorgrids,
        legend cell align=left,
    }}

    \pgfplotsset{tealbar/.style={
        draw=black!50!darkteal,
        fill=darkteal,
    }}

    \pgfplotsset{graybar/.style={
        draw=black!50!gray,
        fill=gray,
    }}

    \pgfplotsset{redbar/.style={
        draw=black!50!darkorange,
        fill=darkorange,
        postaction={
          pattern=north west lines,
        }
    }}

    \centering
    \begin{tikzpicture}
      \begin{axis}[contgraph]

        \addplot[tealbar]   
        table[x=x,y=y]{
          x	y
        1	95.425
        3	76.34
        5	57.26
        7	38.17
        9	19.09
        11	0
        };

        \addplot[redbar]   
        table[x=x,y=y]{
          x	y
        1	95.425
        3	93.119
        5	90.814
        7	88.509
        9	86.203
        11	83.898
        };


      \end{axis}
    \end{tikzpicture}

    \caption{Query Workload Composition}
  \end{subfigure}

  \caption{
    (a) Comparison of \dbl and \basel for different sample sizes used by
    past queries and (b) comparison of \dbl and \basel for different ratios of novel queries in the
    workload.
  }
  \label{fig:exp:contribution}
\end{figure}
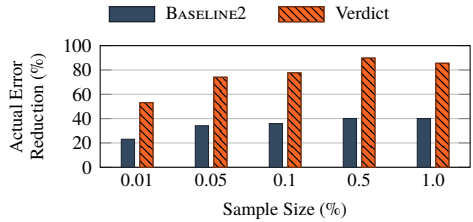
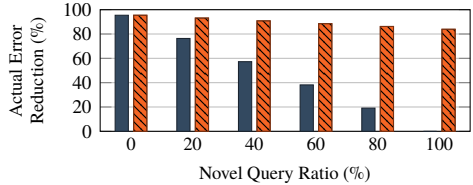

To study the benefits of \dbl's model-based inference, we consider another
system \basel, and make comparisons between \dbl and \basel, using the \tpch
dataset.  \basel is similar to \nol but returns a cached answer if the new query
is identical to one of the past ones.  When there are multiple instances of the
same query, \basel caches the one with the lowest expected error.

\Cref{fig:exp:contribution}(a) reports the average actual error
reductions of \dbl and \basel (over \nol), when different sample sizes were
used for past queries. Here, the same samples were used for new queries. 
The result
shows that both systems' error reductions were large when large sample sizes were used for
the past
queries. However, \dbl consistently achieved higher error reductions
compared to \basel, due to its ability to benefit novel queries as well as
repeated queries (\ie, the queries that have appeared in the past).

\Cref{fig:exp:contribution}(b) compares \dbl and \basel by changing the
ratio of novel queries in the workload.
Understandably, both \dbl and \basel were more effective
for workloads with fewer novel queries (\ie, more repeated
queries); however, \dbl was also
effective for workloads with many novel queries.





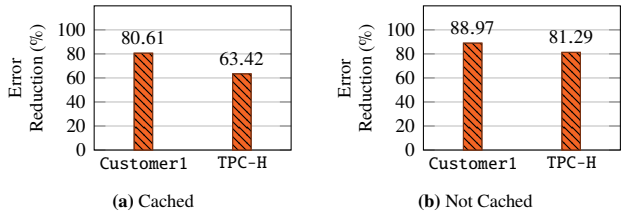
\begin{figure}[t]
  \pgfplotsset{speedupgraph/.style={
      width=45mm,
      height=35mm,
      xtick={1, 2},
      xticklabels={\vertica, \tpch},
      xmin=0.5,
      xmax=2.5,
      ymin=0,
      bar width=7pt,
      xlabel near ticks,
      xlabel near ticks,
      xlabel style={align=center},
      ylabel=Speedup (x),
      legend style={
        at={(0,1.1)},anchor=south west,column sep=2pt,
        draw=none,font=\scriptsize,fill=none,line width=.5pt,
        /tikz/every even column/.append style={column sep=20pt}
      },
      legend columns=2,
      every axis/.append style={font=\scriptsize},
      ybar=10pt,
      area legend,
      xtick style={draw=none},
      x tick label style={yshift=5pt},
      nodes near coords
  }}

  \pgfplotsset{errredgraph/.style={
      width=42mm,
      height=35mm,
      xtick={1, 2},
      xticklabels={\vertica, \tpch},
      xmin=0.5,
      xmax=2.5,
      ymin=0,
      ymax=120,
      ytick={0, 20, ..., 100},
      bar width=7pt,
      xlabel near ticks,
      xlabel shift=-2pt,
      ylabel shift=-2pt,
      xlabel near ticks,
      xlabel style={align=center},
      ylabel style={align=center},
      ylabel=Error\\ Reduction (\%),
      legend style={
        at={(0,1.1)},anchor=south west,column sep=2pt,
        draw=none,font=\scriptsize,fill=none,line width=.5pt,
        /tikz/every even column/.append style={column sep=20pt}
      },
      legend columns=2,
      every axis/.append style={font=\scriptsize},
      ybar=10pt,
      area legend,
      xtick style={draw=none},
      x tick label style={yshift=5pt},
      nodes near coords,
      ymajorgrids,
  }}

  \pgfplotsset{redbar/.style={
      draw=black!50!darkteal,
      fill=darkteal,
  }}

  \pgfplotsset{graybar/.style={
      draw=black!50!darkorange,
      fill=darkorange,
      postaction={
        pattern=north west lines,
      }
  }}

  \begin{subfigure}[b]{0.48\linewidth}
    \begin{tikzpicture}
      \begin{axis}[
          errredgraph,
        ]

        \addplot[graybar]
        table[x=x,y=y]{
          x	y
          1	80.61
          2	63.42
        };



      \end{axis}
    \end{tikzpicture}

    \caption{Cached}
  \end{subfigure}
  ~
  \begin{subfigure}[b]{0.48\linewidth}
    \centering
    \begin{tikzpicture}
      \begin{axis}[
          errredgraph,
        ]

        \addplot[graybar]   
        table[x=x,y=y]{
          x	y
          1	88.97
          2	81.29
        };


      \end{axis}
    \end{tikzpicture}

    \caption{Not Cached}
  \end{subfigure}

  \caption{
    Average error reduction by \dbl (compared to \nol) for the same time budget.
    }
  \label{fig:timebound:errred}
\end{figure}

\subsection{Error Reductions for\\ Time-Bound AQP Engines}
\label{sec:time-bound}
\label{sec:time-bound:setup}
\label{sec:time-bound:results}

Recall that, in \cref{sec:exp}, we demonstrated \dbl's speedup and error reductions over
an online aggregation system. In this section, we show \dbl's error reductions
over a time-bound AQP system. First, we describe our experiment setting.
Next, we present our experiment results.

\ph{Setup}
Here, we describe two systems, \nol and \dbl, which we compare in this section:
\begin{enumerate}[leftmargin=5mm,noitemsep,nolistsep]
\item \nol: This system runs queries on \emph{samples}  of the original
  tables to obtain fast but approximate query answers and their associated
  estimated errors.
  This is the same approach taken by existing AQP engines, such
  as~\cite{surajit-optimized-stratified,aqua1,mozafari_eurosys2013,easy_bound_bootstrap,RusuQT15,mozafari_sigmod2014_demo}.
  Specifically, \nol maintains uniform random samples created offline
  (10\% of the original tables), and it uses the largest samples that are
  small enough to satisfy the requested time bounds.
\item \dbl: This system invokes \nol to obtain raw answers/errors but
 modifies them  to produce improved answers/errors using our proposed inference process.
  \dbl translates the user's requested
  time bound into an appropriate time bound for \nol.
\end{enumerate}
Observe that we are using the term \nol here to indicate a time-bound AQP system in this
section. (In \cref{sec:exp}, we used \nol for an online aggregation system.)

\ph{Experiment Results}
This section presents the error reduction by \dbl compared to \nol. For
experiments, we ran the same set of queries as in \cref{sec:exp} with both \dbl
and \nol described above. For comparison, we used the identical time-bounds for both
\dbl and \nol.
Specifically, we set the time-bounds as 2 seconds and 0.5 seconds for the
\vertica and \tpch datasets cached in memory, respectively; and we set the
time-bounds as 5.0 seconds for both \vertica and \tpch datasets loaded from SSD. 
\Cref{fig:timebound:errred} reports \dbl's error reductions over \nol for each
of four different combinations of a dataset and cache setting.

In \cref{fig:timebound:errred}, one can observe that \dbl achieved large error
reductions (63\%--86\%) over \nol. These results indicate that the users of \dbl
can obtain much more precise answers compared to the users of \nol within the same
time-bounds.



\section{Generalization of Verdict\\ under Data Additions}
\label{sec:append}

Thus far, we have discussed our approach based on the assumption that the database is static, \ie, no
tuples are deleted, added, or updated. In this section, we suggest the possibility of using \dbl
even for the database that allows an important kind of data update: data append.
In \cref{sec:method:appending}, we present the approach, and in
\cref{sec:exp:appending}, we show its effectivess with experiments.

\vspace{5mm}
\subsection{Larger Expected Errors for Old Queries}
\label{sec:method:appending}


A na\"ive strategy to supporting tuples insertions would be to re-execute all past queries every time new tuples
are added to the database to obtain their updated answers.  This solution is
obviously impractical.

Instead, \dbl
still makes use of answers to past queries even when 
new tuples have been added since computing their answers.
The basic idea is to simply lower our confidence in the raw answers of those past queries.

Assume that $q_i$ (whose aggregate function
is on $A_k$) is computed on an old relation
$r$, and a set of new tuples $r^a$ has since been added to
$r$ to form an updated relation $r^u$. Let $\rtans_i^a$ be a
random variable  representing
our knowledge of
$q_i$'s true answer on $r^a$, and $\tans_i^u$ be $q_i$'s true
answer on $r^u$. 

We represent the possible difference  between
$A_k$'s values in $r$ and those in $r^a$ by a random
variable $\shift_k$ with
mean $\mu_k$ and variance $\shiftstd_k^2$. Thus:
$$
\rtans_i^a = \tans_i + \shift_k
$$
The values of $\mu_k$ and  $\shiftstd_k^2$ 
can be estimated using small samples of $r$ and $r^a$. \dbl uses the following lemma to update the
raw answer and raw error for $q_i$:
\begin{lemma}
  \begin{align*}
    E[\rtans_i^u - \rans_i] &= \mu_k \cdot \frac{|r^a| }{|r| + |r^a|}
    \\
    E[(\rtans_i^u - \rans_i - \frac{|r^a| \; \mu_k }{|r| + |r^a|})^2] &= \beta_i^2 +
    \left( \frac{|r^a| \; \shiftstd_k }{|r| + |r^a|} \right)^2
  \end{align*}
where $|r|$ and $|r^a|$ are the number of tuples in $r$ and $r^a$,
respectively.
\end{lemma}

\begin{proof}
Since we represented a snippet answer on the appended relation using a random
variable $\rtans_i^u$, we can also represent a snippet answer on
the updated relation $r^u$ using a random variable. Let the snipept answer on
$r^u$ be $\rtans_i^u$. Then,
\begin{align*}
  E[\rtans_i^u - \rans_i]
  &= E \left[ \frac{|r| \; \tans_i }{|r| + |r^a|}
    + \frac{|r_a| \; \shift_k }{|r| + |r^a|} \right]
    - \tans_i 
  = \frac{|r^a| \; \mu_k}{|r| + |r^a|}
\end{align*}
Also,
\begin{align*}
  &E \left[ (\rtans_i^u - \rans_i -
    \frac{|r^a| \; \mu_k }{|r| + |r^a|})^2 \right] \\
  &= E\left[ \left(
    \tans_i + \frac{|r^a| \; \shift_k }{|r| + |r^a|}
    - \rans_i - \frac{|r^a| \; \mu_k}{|r| + |r^a|}
  \right) \right] \\
  &= E\left[ (\tans_i - \rans_i)^2
    + \left( \frac{|r^a| }{|r| + |r^a|} \right)^2 (\shift_k - \mu_k)^2 \right. \\
   &\qquad \qquad + \left. 2 \left( \frac{|r^a| }{|r| + |r^a|} \right) (\tans_i - \rans_i) (\shift_k - \mu_k)
    \right] \\
  &= \beta_i^2 +
  \left( \frac{|r^a| \; \shiftstd_k }{|r| + |r^a|} \right)^2
\end{align*}
where we used to independence between $(\tans_i - \rans_i)$ and $(\shift_k -
\mu_k)$.
\end{proof}

Once the raw answers and the raw errors of past query snippets are updated using this lemma,
the remaining inference process remains the same.

\begin{figure}[t]

  \pgfplotsset{appending/.style={
      width=65mm,
      height=35mm,
      xmin=1.5,
      xmax=5.5,
      xtick={2, 3, 4, 5},
      xticklabels={5\%, 10\%, 15\%, 20\%},
      ymin=0,
      ymax=5,
      ytick={0, 1, ..., 5},
      yticklabels={0, 1.0, 2.0, 3.0, 4.0, 5.0},
      bar width=5pt,
      ylabel near ticks,
      xlabel near ticks,
      ylabel style={align=center},
      legend style={
        at={(0.0,1.1)},anchor=south west,column sep=0.5pt,
        draw=none,font=\scriptsize,fill=none,line width=.5pt,
        /tikz/every even column/.append style={column sep=-25pt}
      },
      legend columns=2,
      legend cell align=left,
      every axis/.append style={font=\scriptsize},
      ybar,
      area legend,
      xlabel=Proportion of the Number of Appended Tuples,
      x tick style={draw=none},
      x tick label style={yshift=2pt},
      ymajorgrids,
  }}

  \pgfplotsset{graybar/.style={
      draw=black!50!gray,
      fill=gray,
  }}
  \pgfplotsset{darkbar/.style={
      draw=black!50!darkteal,
      fill=darkteal,
  }}
  \pgfplotsset{redbar/.style={
      draw=black!50!darkorange,
      fill=darkorange,
  }}

  \begin{subfigure}[b]{0.9\linewidth}
    \hspace*{7mm}
    \begin{tikzpicture}
      \begin{axis}[
          appending,
          ylabel=Average Error (\%),
          ybar=1pt,
        ]

        \addplot[graybar]   
        table[x=x,y=y]{
          x	y
          1	0.442568
          2	4.373348
          3	4.232032
          4	4.085032
          5	4.086208
        };

        \addplot[bar1]    
        table[x=x,y=y]{
          x	y
          1	0.1521
          2	0.3653
          3	0.1608
          4	0.3547
          5	0.3551
        };

        \addplot[darkbar]   
        table[x=x,y=y]{
          x	y
          1	0.177184
          2	0.412188
          3	0.218736
          4	0.31752
          5	0.196196
        };

        \addlegendimage{empty legend}

        \addplot[redbar]    
        table[x=x,y=y]{
          x	y
          1	0.298116
          2	0.670712
          3	0.839664
          4	1.177568
          5	1.36122
        };

        \addlegendentry{\nol error bound};
        \addlegendentry{\dbl actual error};
        \addlegendentry{\dbl error bound (no adjustment)};
        \addlegendentry{}
        \addlegendentry{\dbl error bound (with adjustment)};

      \end{axis}
    \end{tikzpicture}
  \end{subfigure}

  \vspace{2mm}

  \begin{subfigure}[b]{0.9\linewidth}
    \hspace*{4mm}
    \begin{tikzpicture}
      \begin{axis}[
          appending,
          ylabel=Error bound\\ Violations (\%),
          ymax=100,
          ytick={0, 20, ..., 100},
          yticklabels={0, 20, ..., 100},
          legend style={
            at={(0,1)},anchor=south west,column sep=2pt,
            draw=none,font=\scriptsize,fill=none,line width=.5pt,
            /tikz/every even column/.append style={column sep=15pt}
          },
          legend columns=2,
          bar width=5pt,
        ]


        \addplot[darkbar]   
        table[x=x,y=y]{
          x	y
          2	36.5
          3	40.5
          4	47
          5	79
        };

        \addplot[redbar]    
        table[x=x,y=y]{
          x	y
          2	4
          3	1
          4	3.5
          5	0
        };

        \addlegendentry{no adjustment};
        \addlegendentry{with adjustment};

      \end{axis}
    \end{tikzpicture}
  \end{subfigure}

  \caption{Data append technique (\cref{sec:append}) is highly effective in delivering correct error estimates in face of new data.}
  \label{fig:exp:appending}
  \vspace{\neggap}
\end{figure}
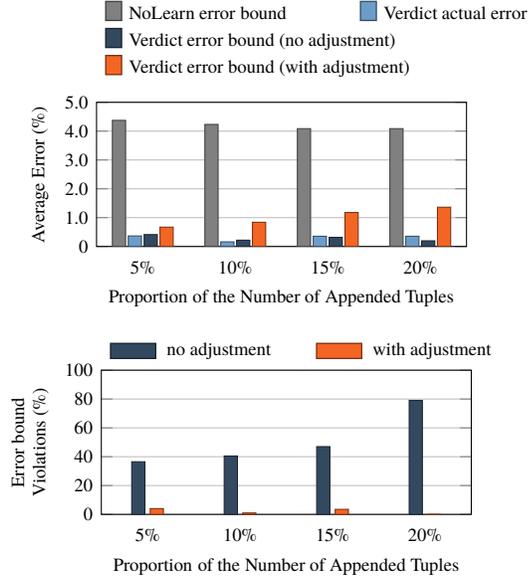

\subsection{Empirical Evaluation for Data Append}
\label{sec:exp:appending}

In this section, we empirically study the impact of new data (i.e., tuple insertions) on \dbl's effectiveness. 
We generated an initial synthetic table with 5 million tuples and appended additional tuples
 to generate different versions of the table. 
 The newly inserted tuples
were generated such that their attribute values gradually diverged from the attribute
values of the original table. 
We distinguish between  these different versions by the ratio of
their newly inserted tuples, e.g., a 5\% appended table means that 250K (5\% of 5 million) tuples were
added. 
We then ran the queries and 
recorded the error bounds of \dblw and \dblwo (our approach with and without the technique introduced in \cref{sec:method:appending}).
We also measured the error bounds of \nol and the actual error.

As shown in \cref{fig:exp:appending}(a), 
\dblwo produced overly-optimistic error bounds (\ie, lower than the actual error) for 15\% and 20\% appends,
whereas \dblw produced valid
error bounds in all cases.
Since this figure shows the \emph{average} error bounds across all queries,
	we also computed the fraction of the individual queries  for which each method's
  error bounds were violated.
In \cref{fig:exp:appending}(b), the Y-axis indicates those cases where the actual error was
	larger than the system-produced error bounds. 
This figure shows more error violations for \dblwo, 
	which increased with the number of new tuples. In contrast, \dblw
produced valid error bounds in most cases, while delivering 
substantial error reductions compared to \nol.

\section{Prevalence of Inter-tuple Covariances in Real-World}
\label{sec:covariance:study}

\afterrev{
In this section, we demonstrate the existence of
the inter-tuple covariances in many real-world datasets by analyzing well-known datasets from 
the UCI repository~\cite{Lichman:2013}. We analyzed the following well-known 16 datasets:
 cancer, glass, haberman, ionosphere, iris, mammographic-masses, optdigits, parkinsons,
pima-indians-diabetes, segmentation, spambase, steel-plates-faults, transfusion, vehicle,
vertebral-column, and yeast.

We extracted numeric attributes (or equivalently, columns) from those datasets and composed each of
the datasets into a relational table. Suppose a dataset has $m$ attributes. Then, we computed the correlation
between adjacent attribute values in the $i$-th column when the column is sorted in order of another
$j$-th column---$i$ and $j$ are the values (inclusively) between 1 and $m$, and $i \ne j$. Note that
there are $m (m-1) / 2$ number of pairs of attributes for a dataset with $m$ attributes. We analyzed
all of those pairs for each of 16 datasets listed above.}

\begin{figure}[t]
  \pgfplotsset{laterr/.style={
      width=70mm,
      height=32mm,
      ybar interval,
      ylabel near ticks,
      ylabel shift=-1pt,
      xlabel near ticks,
      xlabel shift=-3pt,
      ylabel style={align=center},
      xlabel=(Normalized) Inter-tuple Covariance,
      ylabel=Percentage (\%),
      legend style={
        at={(1,1)},anchor=north east,column sep=2pt,
        draw=none,fill=none,font=\tiny,line width=.5pt,
        /tikz/every even column/.append style={column sep=20pt}
      },
      legend columns=1,
      every axis/.append style={font=\scriptsize},
      x tick label style={xshift=7pt,anchor=east,rotate=60},
      ymajorgrids,
      clip=false,
  }}

  \pgfplotsset{graydot/.style={
      draw=black!50!darkteal,
      fill=darkteal,
  }}

  \centering
  \begin{tikzpicture}
    \begin{axis}[
        laterr,
        xmin=0.5,
        xmax=13.5,
        xtick={1, 2, ..., 13},
        xticklabels={-0.1, 0.0, 0.1, 0.2, 0.3, 0.4, 0.5, 0.6, 0.7, 0.8, 0.9, 1.0},
        ymin=0,
        ymax=60,
        ytick={0, 20, ..., 60},
      ]
      \addplot[graydot] plot coordinates
      {
        (1, 0.02)
        (2, 2.05)
        (3, 2.12)
        (4, 2.36)
        (5, 2.29)
        (6, 3.25)
        (7, 4.63)
        (8, 6.34)
        (9, 7.09)
        (10, 10.05)
        (11, 13.25)
        (12, 46.55)
        (13, 0)
      };

      \node[rotate=60,anchor=east,font=\scriptsize] at (axis cs: 1.0, -7) {-0.2};

    \end{axis}
  \end{tikzpicture}

  \vspace{\neggap}
  \caption{\emph{Inter-tuple Covariances} for 16 real-life UCI datasets.
  }
  \label{fig:overview:corr}
\end{figure}
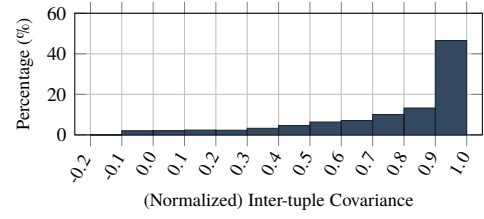

\afterrev{
\Cref{fig:overview:corr} shows the results of our analysis. The figure reports the percentage of
different levels of correlations (or equivalently, normalized inter-tuple covariances) between
adjacent attributes. One can observe that there existed strong correlations in the datasets we
analyzed. Remember that the users of \dbl do not need to provide any information regarding the
inter-tuple covariances; \dbl automatically detects them as described in
\cref{sec:learn}, relying on
the past snippet answers stored in the query synopsis.}

\section{Technical Details}
\label{sec:tech_details}

In this section, we present the mathematical details we have omitted in the
main body of this paper. First, we describe the analytics expression for the
double-integrals in \cref{eq:simple_cov}. Second, we extend the result of
\cref{sec:model} to categorical attributes. Third, we provides details on some
correlation parameter computations.

\vspace{2mm}
\subsection{Double-integration of Exp Function}
\label{sec:double_int}

\rev{C2.1c}{
For the analytic computation of \cref{eq:simple_cov}, we must be able to
analytically express the result of the following double integral:
\begin{align}
  f(x, y) = \int_a^b
\int_c^d \,
\exp \left( -\frac{(x-y)^2}{z^2} \right)
\, d x \, d y
\label{eq:double_int}
\end{align}
To obtain its indefinite integral, we used
the Symbolic Math Toolbox in Matlab, which yielded:
\[
  f(x, y) = - \frac{1}{2} \left(
  z^2 \exp \left( - \frac{(x-y)^2}{z^2} \right)
  \right)
  - \frac{\sqrt{\pi}}{2} z (x-y) \erf \left( \frac{x-y}{z} \right)
\]
where $\erf(\cdot)$ is the \emph{error function}, available from most
mathematics
libraries. Then, the definite integral in \cref{eq:double_int} is obtained by
$f(b, d) - f(b, c) - f(a, d) + f(a, c)$.
}

\subsection{Handling Categorical Attributes}
\label{sec:discrete}

\rev{C2.2}{
Thus far, we have assumed that all dimension attributes are numeric. This
section describes how to handle dimension attributes that contain both numeric
and categorical attributes.
Let a tuple $\mtuple = (a_1, \ldots, a_c, a_{c+1}, \ldots, a_l)$, where $c$ is
the number of categorical attributes; the number of numeric attributes is
$l - c$. Also $\mtuple_c = (a_1, \ldots, a_c)$ and $\mtuple_l = (a_{c+1},
\ldots, a_l)$. The covariance between two query snippet answers in \cref{eq:query_cov}
is extended as:
\[
  \sum_{\mtuple_c \in \filter_{i}^{c}}
  \sum_{\mtuple'_c \in \filter_{j}^c}
  \int_{\mtuple_l \in \filter_{i}^l} \int_{\mtuple_l' \in \filter_{j}^l}
    \cov(\nu_g(\mtuple), \nu_g(\mtuple')) \; d\mtuple \; d\mtuple'
\]
where $\filter_{i}^c$ is the set of $\mtuple_c$ that satisfies $q_i$'s selection
predicates on categorical attributes, and $\filter_{i}^l$ is the set of
$\mtuple_l$ that satisfies $q_i$'s selection predicates on numeric attributes.
The first question, then, is how to define the inter-tuple covariance, \ie,
$\cov(\nu_g(\mtuple), \nu_g(\mtuple'))$, when two arbitrary tuples $\mtuple$ and
$\mtuple'$ follow the schema with both categorical and numeric
attributes. For this, \dbl extends the previous inter-tuple covariance in
\cref{eq:tuple_cov}, which was defined only for numeric attributes, as follows:
\begin{align*}
  &\cov(\nu_g(\mtuple), \nu_g(\mtuple'))
  \approx
  \sigma_{g}^2 \cdot 
  \prod_{k=1}^c 
    \delta(\tcol_k, \tcol'_k)
  \prod_{k=c+1}^{l} \exp \left(
    -\frac{(\tcol_k - \tcol'_k)^2}{l_{g,k}^2} \right)
\end{align*}
where $\delta(\tcol, \tcol')$ returns 1 if $\tcol = \tcol'$ and 0
otherwise. The inter-tuple covariance between two tuples become zero
if they include different categorical attribute values. Note that this is a
natural choice, since 
the covariance between the two random variables, independently and identically drawn from the same distribution, is zero.
With the above definition
of the inter-tuple covariance, $\cov(\rtans_i, \rtans_j)$ is expressed as:
\begin{align}
  \begin{split}
  &\sigma_{g}^2
  \prod_{k=1}^{c}
    |F_{i,k} \cap F_{j,k} | \\
  &\prod_{k=c+1}^{l}
    \int_{s_{i,k}}^{e_{i,k}}
    \int_{s_{j,k}}^{e_{j,k}} \,
    \exp \left( -\frac{(\tcol_k-\tcol'_k)^2}{l_{g,k}^2} \right) \,
    d a'_k a_k
  \end{split}
  \label{eq:cov_exact}
\end{align}
where $F_{i,k}$ and $F_{j,k}$ are the set of the $A_k$'s categorical attribute
values used for the \texttt{in} operator in $i$-th and $j$-th query snippet, respectively.
If $q_i$ includes a single equality constraint for a categorical attribute
$A_k$, \eg, \texttt{$A_k$ = 1}, the equality constraint is conceptually treated as 
the \texttt{in} operator with the list including only that
particular attribute value. If no constraints are specified in $q_i$ for a categorical attribute
$A_k$, $F_{i,k}$ is conceptually treated as a universal set including all
attribute values in $A_k$.
The above expression can be computed efficiently, since
counting the number of common elements between two sets can be performed in a
linear time using a hash set, and the double integral of an exponential
function can be computed analytically (\cref{sec:double_int}).
}

\subsection{Analytically Computed Parameter Values}
\label{sec:anal_param}

\rev{C2.4}{
While \dbl learns correlation parameters, \ie, $l_{g,1}, \ldots, l_{g,l}$,
by solving an optimization
problem, the two other parameters, \ie, the expected values of query answers
(namely $\qmeanvec$) and the multiplier for $\rho_g(\mtuple, \mtuple')$ (namely $\sigma_g$),
are analytically computed as follows. Recall that $\qmeanvec$ is
used for computing model-based answers and errors (\cref{eq:infer2}), and
$\sigma_g$ is used for computing the covariances between pairs of query answers
(\cref{eq:cov_exact}).

First, we use a single scalar value $\qmean$ for the expected values of the prior
distribution; that is, every element of $\qmeanvec$ is set to $\qmean$ once we
obtain the value. Note that it only serves as the
means in the prior distribution.  We take different
approaches for \texttt{AVG($A_k$)} and \texttt{FREQ(*)} as follows. For \texttt{AVG(*)}, we
simply set $\qmean = \sum_{i=1}^n \estans_i / n$; whereas, for \texttt{FREQ(*)},
we set $\qmean = \sum_{i=1}^n \estans_i /
|\filter_i|$ where $|\filter_i|$ is the area of the hyper-rectangle
$\prod_{k=c+1}^l (s_{i,k},e_{i,k})$ specified as $q_i$'s selection
predicates on numeric attributes.

Second, observe that $\sigma_g^2$ is equivalent to the variance of
$\nu_g(\mtuple)$. For \texttt{AVG($A_k$)}, we use the variance of $\estans_1,
\ldots, \estans_n$; for \texttt{FREQ(*)}, we use the variance of $\estans_1 /
|\filter_i|, \ldots, \estans_n / |\filter_i|$.
We attempted to learn the optimal value for $\sigma_g^2$ in the course of solving the optimization
problem (\cref{eq:log_likelihood}); however, the local optimum did not produce a
model close to the true distribution.
}

\ignore{
\subsection{Proof to \cref{thm:accuracy} and \cref{lemma:time}}
\label{sec:theorem_proof}

}

\ignore{
\begin{figure}[t]
  \centering

  \pgfplotsset{axisIntroFigure/.style={
          width=48mm,
          height=30mm,
          xmin=0,
          xmax=1,
          ymin=0,
          ymax=6,
          legend style={at={(-0.06,1.24)},anchor=south west,column sep=1.5pt,
          draw=none,font=\scriptsize,fill=none,line width=.5pt},
          legend cell align=left,
          legend columns=1,
          clip=false,
          ytick={0, 2, 4, 6},
          xtick={0, .2, .4, .6, .8, 1.0},
          ylabel=$\nu_g(\mtuple)$,
          xlabel=$\tcol_d$,
          ylabel near ticks,
          xlabel near ticks,
          xlabel shift=-1mm,
  }}

  \pgfplotsset{curve/.style={%
          mark=none,
          opacity=1.0
  }}

  \tikzset{barstyle/.style={%
          opacity=0.3,
          fill=darkteal,
  }}

  \tikzset{legendnode/.style={%
          opacity=0.7,
          fill=darkteal,
          minimum width=5mm,
          minimum height=2.5mm,
          anchor=south west,
  }}

  \begin{subfigure}{0.48\linewidth}
    \centering
    \begin{tikzpicture}
      \begin{axis}[axisIntroFigure]

        \addplot[curve,thick,darkteal] table [x=x, y=y, col sep=comma] {data/lowFreqData.csv};            
        \addplot[curve,tomato,ultra thick] table [x=x, y=y, col sep=tab] {data/lowFreqModel.csv};    

        \addlegendentry{Groundtruth}

        \draw [barstyle] (axis cs:0.658,0) rectangle (axis cs:0.799,2.639);
        \draw [barstyle] (axis cs:0.187,0) rectangle (axis cs:0.32,2.544);
        \draw [barstyle] (axis cs:0.871,0) rectangle (axis cs:0.972,0.714);
        \draw [barstyle] (axis cs:0.867,0) rectangle (axis cs:1.0,0.844);
        \draw [barstyle] (axis cs:0.852,0) rectangle (axis cs:1.0,0.843);
        \draw [barstyle] (axis cs:0.357,0) rectangle (axis cs:0.492,2.36);
        \draw [barstyle] (axis cs:0.265,0) rectangle (axis cs:0.415,2.927);
        \draw [barstyle] (axis cs:0.104,0) rectangle (axis cs:0.281,2.497);
        \draw [barstyle] (axis cs:0.594,0) rectangle (axis cs:0.71,3.88);
        \draw [barstyle] (axis cs:0.34 ,0) rectangle (axis cs:0.454,2.655);

        \node [legendnode,darkteal,opacity=0.5] at (axis cs:0.02, 6.5)
        [label={[font=\scriptsize]east:Queries}] {};

      \end{axis}
    \end{tikzpicture}
    \vspace{-5mm}
    \caption{Simple groundtruth}
  \end{subfigure}
  ~
  \begin{subfigure}{0.48\linewidth}
    \centering
    \begin{tikzpicture}
      \begin{axis}[axisIntroFigure,clip=true]

        \addplot[curve,darkteal] table [x=x, y=y, col sep=comma,forget plot] {data/mixedFreqData.csv};            
        \addplot[curve,tomato,ultra thick] table [x=x, y=y, col sep=tab] {data/mixedFreqModel.csv};    

        \addlegendentry{Internal Model}

        \draw [barstyle] (axis cs:0.658,0) rectangle (axis cs:0.799,2.621);
        \draw [barstyle] (axis cs:0.187,0) rectangle (axis cs:0.32,2.706);
        \draw [barstyle] (axis cs:0.871,0) rectangle (axis cs:0.972,0.614);
        \draw [barstyle] (axis cs:0.867,0) rectangle (axis cs:1.0,0.762);
        \draw [barstyle] (axis cs:0.852,0) rectangle (axis cs:1.0,0.795);
        \draw [barstyle] (axis cs:0.357,0) rectangle (axis cs:0.492,2.339);
        \draw [barstyle] (axis cs:0.265,0) rectangle (axis cs:0.415,2.795);
        \draw [barstyle] (axis cs:0.104,0) rectangle (axis cs:0.281,2.731);
        \draw [barstyle] (axis cs:0.594,0) rectangle (axis cs:0.71,3.619);
        \draw [barstyle] (axis cs:0.34 ,0) rectangle (axis cs:0.454,2.532);

      \end{axis}
    \end{tikzpicture}
    \vspace{-5mm}
    \caption{Complex groundtruth}
  \end{subfigure}

  \caption{\dbl's optimization process finds the simplest model that
      explains the past queries and their answers. This process effectively ignores the noisy components (in
    Figure (b)) that do not affect the query answers.}
  \label{fig:learn:simplest}
  \vspace{\neggap}
\end{figure}

Informally, \dbl finds the \emph{simplest model that is consistent with the raw
answers (and raw errors) to past queries.} To illustrate such \dbl's behavior,
we prepared two types of datasets: in the first dataset
(Figure~\ref{fig:learn:simplest}(a)), the aggregate (represented on Y-axis) is a
slowly changing function of $\tcol_d$, where $\tcol_d$ is the value for one of dimension
attributes; while in the second dataset
(Figure~\ref{fig:learn:simplest}(b)), the aggregate is a fast changing function
of $\tcol_d$. One can observe that the models determined by \dbl are identical for
those two datasets even if those two aggregates were different, which is because
the answers to the past queries (represented in grey boxes) are the same, and
the only information that \dbl relies on is the answers (and errors) to the
past queries. Although the internal model does not truly reflect the groundtruth
in the second dataset, the model would be still useful for computing improved
answers because the high frequency components barely affects query answers when
aggregated (they cancel out to zero).
}


\end{document}